\documentclass[prd,amssymb,aps,twocolumn]{revtex4}

\usepackage{graphicx}
\usepackage{subfigure}

\begin{document}

\title{Scalar field confinement as a model for accreting systems}
\date{\today}
\author{M. Megevand$^{1}$, I. Olabarrieta$^{1,2}$, L.~Lehner$^{1}$}
\affiliation{$1$ Department of Physics and
Astronomy, Louisiana State University, 202 Nicholson Hall, Baton
Rouge, Louisiana 70803-4001, USA\\
$2$ TELECOM Unit, ROBOTIKER-Tecnalia, Ed. 202 Parque Tecnol\'ogico
Zamudio E-48170, Bizkaia, SPAIN}

\begin{abstract}
We investigate the possibility to localize scalar field configurations
as a model for black hole accretion. We analyze and resolve difficulties 
encountered when localizing scalar fields in General Relativity.
We illustrate this ability with a simple spherically symmetric model
which can be used to study features of accreting shells around a black hole.
This is accomplished by prescribing a scalar field with a coordinate
dependent potential. Numerical solutions to the Einstein-Klein-Gordon equations
are shown, where a scalar filed is indeed confined within a region surrounding a
black hole. The resulting spacetime can be described in terms of simple harmonic 
time dependence.
\end{abstract}

%---Load Packages---------------------------------
%\usepackage{amssymb}
%\usepackage{graphicx}
%\usepackage[draft]{graphicx}  % display graph box and name only

%---Aliases---------------------------------------
\newcommand{\be}{\begin{equation}}
\newcommand{\ee}{\end{equation}}
\newcommand{\bea}{\begin{eqnarray}}
\newcommand{\eea}{\end{eqnarray}}
\newcommand{\bw}{ \begin{widetext} }
\newcommand{\ew}{ \end{widetext} }
\renewcommand{\d}{\partial}
\newcommand{\eq }[1]{(\ref{#1})}
\newcommand{\cov}{\nabla}
\newcommand{\half}{\frac{1}{2}}

\newcommand{\ph}{\varphi}
\newcommand{\DS}{\displaystyle}
% Sub and super indexes not aligned vertically:
\newcommand{\T}[1]{ {T^{(#1)}} }

\renewcommand{\aa}{ \tilde{\alpha} }

%---Body------------------------------------------
%\begin{document}
\maketitle
%\begin{abstract}
%\end{abstract}

%%%%%%%%%%%%%%%%%%%%%%%%%%%%%%%%%%%%%%%%%%%%%%%%%%
\section{Introduction}
%%%%%%%%%%%%%%%%%%%%%%%%%%%%%%%%%%%%%%%%%%%%%%%%%%
Self-gravitating scalar field configurations
have been very useful in many aspects of gravitational theory.
Their role as describing matter models (eg.\cite{MTW,scheelteuk,kaup,ruffini,seidelsuen});
as governing mechanisms to model inflationary scenarios (eg. \cite{liddle,Lidsey:1995np});
as probes of strong curvature regions (eg.\cite{choptuik,Gundlach:1997wm}), etc. has made them an
ideal tool in a number of fronts. In this work we examine exploiting
scalar fields to mimic some salient properties of accreting black hole
systems. To this end, it is desirable to explore a configuration
where the scalar field simulates an accretion disk
surrounding a black hole. For this purpose, one should be able to confine
the scalar field within some compact region surrounding the black
hole.
Since massless scalar fields radiate away to infinity, the model sought after
should include a mechanism that will prevent this from happening, at least
to some non-trivial extent. (The existence of bound states for particular
cases in spherical symmetry are studied in \cite{bound,bound2}).

One way to confine the scalar field would be by employing
a potential well which would introduce some sort of barrier and thus allow
for confinement.  The use of carefully chosen potentials is common practice
with scalar fields, and are usually functions of the field itself. 
Examples of this kind of potential are the quadratic ($V(\phi)\propto\phi^2$) 
--that introduces a mass term--;
and the quartic ($V(\phi)\propto\phi^4$). However, that kind of
potential does not allow for confining the scalar field 
within a specific region of space, that one can specify {\em a priori.}

What we are looking for is a potential that somehow
depends on the coordinates and in particular can be chosen to describe
a potential well within a region. However, this proposition seems a priori at odds
with maintaining covariance. 
The difficulty one encounters with a coordinate-dependent
potential, is that the corresponding stress-energy tensor is
in general inconsistent, in the sense that its divergence will not be zero
for a non-trivial scalar field. This fact, together with the Einstein
equations, would imply
that the Bianchi identities are not satisfied.

Faced with this situation a possible way of confining the scalar field would
be to introduce a background with respect to which coordinates could be
defined. This approach would be in line with bi-metric 
theories of gravity (eg.\cite{rosen}). Another approach would be to fix a suitable
coordinates already at the level of the action as is done in \cite{kuchar} through
some suitably introduced Lagrande multipliers. This procedure then provides a way
to convariantly adopt coordinates which could, in principle, be used in the potential.
However, this method would strongly link the adopted gauge with the type of 
potential introduced and it is not yet clear whether it can be made of practical use. 
An alternative way, which is the one we pursue here, 
is to exploit symmetry   considerations without resorting to introducing any 
other feature in the problem. The existence of the symmetry provides a simple
way to consistently introduce a coordinate-dependent potential in the problem.
Certainly, while more restricted than other possible viable options, this approach
is the more direct one. 
A particular case of a coordinate dependent potential has already been
implemented in~\cite{frans,unpub} to effectively simulate angular
momentum in spherical or axial symmetry. 

In this work we concentrate mainly on the case of spherical symmetry, but
give prescriptions for the implementation of potentials in both
spherical and axial symmetry. We will see that, if the space is
spherically symmetric, we can implement a potential that depends on the
areal radius. In the same way, for an axially symmetric spacetime, the 
potential can depend on the length of the closed integral curves defined
by the associated killing vector.
Even though one will not be able to specify the potential as
an arbitrary function of any coordinate, 
one may still be
able to confine a scalar field to some region, as is it shown in this
work for the case of spherical symmetry. This fact will become
apparent in section \ref{sphericalcase} and in its applications in the
rest of this work.

This paper is organized as follows. In section~\ref{derivations}, we
study the specification of a 
stress-energy tensor for a scalar field with a coordinate dependent
potential. Showing that such implementation is possible when the
space-time possesses a symmetry. In particular, the case of spherical
symmetry is studied in depth (we also consider an
axi-symmetric case in an appendix). In section~\ref{TheEquations} we
describe the formulation used, and the resulting equations. In
section~\ref{numerics} we discuss how the equations are solved
numerically, after obtaining initial data by two different methods. 
In section~\ref{results} we show and analyze the
numerical solutions obtained, finding that, after some transient behavior, 
the scalar field reaches a state described by a simple harmonic time dependence
and remains
confined to a region surrounding the black hole. We have confirmed these
for initial masses of the scalar field up to $50\%$ of that of the
black hole. Finally, we make some final remarks in section~\ref{conclusions}.
In all this work we use Einstein's index notation and geometrized units.

%%%%%%%%%%%%%%%%%%%%%%%%%%%%%%%%%%%%%%%%%%%%%%%%%%
\section{Scalar Field on a Coordinate-Dependent Potential} \label{derivations}
%%%%%%%%%%%%%%%%%%%%%%%%%%%%%%%%%%%%%%%%%%%%%%%%%%
In this section we study the specification of a stress-energy tensor for a
scalar field with a coordinate dependent potential. Our motivation is
to somehow confine a scalar field within a region around a 
black hole. The resulting system would share features of a
black hole interacting with an accretion disk. We will see
that this can be done when the space-time posses a symmetry. However,
the specification of such potential is not completely arbitrary since
it must depend on the coordinates only through
some particular function. 
Knowing the approximate dependence of that function on the
coordinates, one can then construct a potential that  confines the
scalar field.  

Before presenting our approach, we include an overview of how
the equations of motion are obtained  
from a stress-energy tensor in the case of a coordinate-independent
potential. Then, based on that procedure, we will study the generalization
to the case of a coordinate-dependent potential.

The equations of motion for a real scalar field $\phi$ on a
coordinate-independent potential can be derived from the stress-energy
tensor  
\be
  T_{ab} = {T^{(k)}}_{ab} + {T^{(p)}}_{ab},
\ee
where, for later convenience, we have split this tensor into what we
call the ``kinetic'' and  ``potential'' terms:
\be  \label{T0}
  {T^{(k)}}_{ab} \equiv (\cov_a\phi) (\cov_b\phi) - \half g_{ab}(\cov_c\phi)
  (\cov^c\phi), 
\ee
\be
   {T^{(p)}}_{ab} \equiv - \half g_{ab} V(\phi). \label{T1_old}
\ee
The kinetic part, ${T^{(k)}}_{ab}$, corresponds  to a  massless scalar field
without a potential. 

The equations of motion can be obtained \cite{wald,MTW} through the condition
\be
  \cov_a {T^a}_b = 0 \;,
  \label{divergence}
\ee
which must be satisfied to be consistent with a 
covariant theory. Equation \eq{divergence}
can be re-expressed with $\cov_b \phi$ as a 
common factor,
\be
  0 = \cov_a {T^a}_b = (\cov_b \phi) \; \mathcal{L}(\phi),
\label{common_factor}
\ee
where $\mathcal{L}(\phi)$ contains second order derivatives of $\phi$.
The equations of motion for a non-trivial scalar field is then
\be \mathcal{L}(\phi) = 0 \label{Lis0}    \,. \ee
For example, for $V(\phi)=m^2\phi^2$ we obtain the Klein-Gordon equation,
\be
  \mathcal{L}(\phi) \equiv \left( \cov_a \cov^a - m^2 \right) \phi=0\;.
\ee
This is analogous to the Lagrangian approach, where the variation of the
action is set to zero, and, after integrating by parts, the integrand becomes
$\delta\phi\mathcal{L}(\phi)$.  \\

After this detour, we now turn our attention back to the case of interest, 
the implementation of a coordinate-dependent potential. Our discussion is based
on the precedent one though now generalizing it to the case of 
a coordinate-dependent potential $V(x^c,\phi)$. 

A naive first approach would be to replace occurrences of $V(\phi)$ 
in \eq{T1_old} by $V(x^c,\phi)$. However, this will bring an unfortunate
consequence, namely that one can now no longer express the divergence 
of ${T^a}_b$ in the form given by eqn \eq{common_factor},
where $\cov_b \phi$ appears as a common factor. Instead one has
\bea
0=  \cov_a {T^a}_b  &=&  (\cov_b \phi) \left(\cov_a \cov^a \phi -\half \frac{\d
    }{\d \phi}V(x^c,\phi)\right) \nonumber \\
   && - \half \frac{\d }{\d x^b}V(x^c,\phi). \label{not_common_factor}
\eea

The crucial difference with eqn. \eq{common_factor} is that several
(independent) equations must be satisfied by the real scalar field
$\phi$. As a result, the system of equations will be generically inconsistent.

To resolve this problem we start by: (i) adopting a different
ansatz for ${T^{(p)}}_{ab}$ (equation \eq{T1} below), and (ii)
imposing symmetry conditions on the scalar field.

First, consider setting ${T^{(p)}}_{ab}$, instead of being given by equation \eq{T1_old},
to be the product of a function of $\phi$ and a coordinate
dependent tensor,
\be 
  \T{p}_{ab} \equiv H_{ab}(x^c)\;f(\phi) \label{T1},
\ee
where the function $f$ is independent of $x^c$ and the tensor $H_{ab}$ is
independent of $\phi$.
Now, find a suitable $H_{ab}$ such that $\cov_a {T^a}_b$ takes the
form of equation \eq{common_factor},
this will induce conditions on $H_{ab}$. Under this choice the divergence  of the
stress-energy tensor results
\be
  \cov_a {T^a}_b = (\cov_b \phi) \cov_a \cov^a \phi +
  \frac{\d f}{\d \phi}(\cov_a \phi)
  \;{H^a}_b + f(\phi) \;\cov_a {H^a}_b \;. \label{div_T}
\ee
Now, we look for conditions that would allow us to 
express the \emph{r.h.s.} of equation \eq{div_T} in such
a way that $\cov_b \phi$ appears as a common factor. 
Since ${H^a}_b$ is independent of $\phi$, $\cov_b\phi$ cannot appear
in the last term of \eq{div_T}; Then, that term must be zero, resulting in
the first condition on ${H^a}_b$, 
\be
  \cov_a {H^a}_b = 0 \label{div_H}\;.
\ee
We now consider the second term in the \emph{r.h.s.}; the condition
\be
(\cov_a\phi)\; {H^a}_b = (\cov_b\phi)\;h(x^c)\;, \label{ab}
\ee
for some scalar $h(x^c)$, ensures that that term has $\cov_b\phi$  as a common factor.
Equation~\eq{ab} is satisfied for any scalar field $\phi$ if 
\be
  {H^a}_b = h(x^c){\delta^a}_b\;. \label{scalar}
\ee
However, this condition, together with equation \eq{div_H},
implies that $h(x^c)$ is a constant. This means that ${T^{(p)}}_{ab}$ is
of the form~\eq{T1_old} (with $V$ independent of $x^c$). Thus, for an arbitrary
scalar field, and without any further structure in the spacetime, space-dependent
potentials can not be considered. However, by imposing further conditions
on the scalar field $\phi$, ${H^a}_b$ can indeed be chosen with further structure than 
that of equation~\eq{scalar} while still satisfying equation \eq{ab}. To this end, we
consider\footnote{This equation can be thought just as the
  definition of the tensor ${A^a}_b$} the tensor ${H^a}_b$ of the form
\be
 {H^a}_b = h(x){\delta^a}_b + {A^a}_b \label{Adef}\;.
\ee
Replacing \eq{Adef} into \eq{ab} we find
\be
 (\cov_a \phi) {A^a}_b = 0\;. \label{fA}
\ee
The simplest case is the one with ${A^a}_b=0$ for which ${H^a}_b$ is given by
\eq{scalar}. More general cases arise when
$\phi$ is independent on one of the coordinates, lets say $\d_{x^3}
\phi\equiv\cov_3\phi=0$. Here one can adopt ${A^3}_3$
arbitrarily and set all other components to zero, thus satisfying
equation \eq{fA}. 

In this particular case, ${H^a}_b$ takes the form
\be
 {H^a}_b = \left[\begin{array}{cccc}
                  h & 0 & 0 & 0 \\
		  0 & h & 0 & 0 \\
		  0 & 0 & h & 0 \\
		  0 & 0 & 0 & b 
                  \end{array}\right]   \label{one}
\ee
for some functions $h(x^c)$ and $b(x^c)$.

Similarly, when $\phi$ does not depend on two of
the coordinates, lets say $\d_{x^2}
\phi=0$, $\d_{x^3}\phi=0$, one can choose 
\be
 {H^a}_b = \left[\begin{array}{cccc}
                  h & 0 & 0 & 0 \\
		  0 & h & 0 & 0 \\
		  0 & 0 & b & 0 \\
		  0 & 0 & 0 & c 
                  \end{array}\right] .  \label{two}
\ee
Analogous results are obtained when some of its derivatives are linearly 
related. For example, if $\d_{x^3}\phi=c\d_{x^2}\phi$, one can
adopt ${A^3}_3$ arbitrarily and set ${A^2}_3=-c{A^3}_3$ keeping all other components
zero. With this choice, equation \eq{fA} will be satisfied and ${H^a}_b$ will then be given in
terms of two functions $h(x^c)$ and $b(x^c)$ in a slightly different way
as is \eq{one}. \\

Summarizing, we have seen that a coordinate dependent potential can be
implemented if the following conditions are satisfied: (i) its derivatives are 
linearly dependent (this includes the possibility of one or more of
them being zero). (ii) The ``potential'' part of the stress-energy tensor is
given by \eq{T1}, with ${H^a}_b$ satisfying $\cov_a{H^a}_b=0$ and
being expressible in the form \eq{one}, \eq{two}, or similar expressions depending
on how condition (i) is fulfilled.

In the next section we will consider in detail the case of spherical
symmetry. 

%%%%%%%%%%%%%%%%%%%%%%%%%%%%%%%%%%%%%%%%%%%%%%%%%%
\subsection{Spherical Symmetry} \label{sphericalcase}
%%%%%%%%%%%%%%%%%%%%%%%%%%%%%%%%%%%%%%%%%%%%%%%%%%

We will now concentrate on the case of spherical symmetry. The line element
can be written in the form
\be
  ds^2 = -N^2 dt^2 + g_{rr}(dr+\beta dt)^2 + g_{\Omega}
  d\Omega^2,
\ee
where $N$, $g_{rr}$, $\beta$, and $g_{\Omega}$ are functions of $t$ and $r$.
We assume that we can adopt coordinates so that $\d_\theta\phi=\d_\ph\phi=0$. 
Then, ${H^a}_b$ is given by \eq{two},
 with the additional condition that $b=c$ due to the
spherical symmetry. ${H^a}_b$ is then
\be  \label{sH}
 {H^a}_b = \left[\begin{array}{cccc}
                  h & 0 & 0 & 0 \\
		  0 & h & 0 & 0 \\
		  0 & 0 & b & 0 \\
		  0 & 0 & 0 & b 
                  \end{array}\right]  ,
\ee
with $h$ and $b$ functions of $t$ and $r$.

The evaluation of $\cov_a {H^a}_b$ gives rise to non-trivial equations
only on the $t$ and $r$ components,
\bea   
      \frac{dg_{\Omega} }{ dt\;\;}(h-b)   \label{hb_t}
 +2 g_{\Omega}   \frac{dh}{dt}   &=& 0, \\
       \frac{dg_{\Omega} }{ dr\;\;}(h-b)  \label{hb_r}
 +2 g_{\Omega}   \frac{dh}{dr}   &=& 0.
\eea

In order to obtain a family of solutions to these equations we will
demand that $h$ depends on the coordinates only
through $g_{\Omega}$: $h(t,r)=h(g_{\Omega}(t,r))$.
With this condition, we have that
\be
 \frac{dh}{dx^i} = \frac{\d h}{\d g_{\Omega}} 
                   \; \frac{dg_{\Omega}}{dx^i} 
\ee
for $x^i=(t,r)$. Substituting this into either equation \eq{hb_t} or  \eq{hb_r}, we
obtain an expression for $b$ in terms of $h$,
\be \label{bh}
 b = h + g_{\Omega} \frac{\d h}{\d g_{\Omega}} .
\ee

We have just seen that, if $h$ depends on the coordinates only
through $g_{\Omega}$,
and $b$ is given in terms of $h$ by \eq{bh}, the prescription \eq{sH}
for the tensor ${H^a}_b$ allows us to express $\cov_a {T^a}_b$
with $\cov_b \phi$ as a common factor. More explicitly:
\be  \label{shperical_cf}
 \cov_a {T^a}_b =  (\cov_b\phi)\left(\cov_a\cov^a\phi + 
 \frac{\d f}{\d \phi} h(g_{\Omega}) \right).
\ee
Notice that, if one wanted to calculate $\cov_a {T^a}_b$ without
setting $\d_\theta \phi = \d_\ph \phi = 0$ at the onset, one
would obtain \eq{shperical_cf}, but with 
 $h(g_{\Omega})$ replaced by $b(g_{\Omega})$ for the
angular components $\cov_a {T^a}_\theta$ and $\cov_a
{T^a}_\ph$. However, because those terms are actually multiplied by zero,
equation~\eq{shperical_cf} is true for all four components.

Setting the {\em r.h.s} of \eq{shperical_cf} to zero we obtain the
equation of motion for $\phi$,
\be
 \cov_a\cov^a\phi + 
 \frac{\d f}{\d \phi} h(g_{\Omega}) = 0,
\ee
where we remind the reader that $f$ is an arbitrary function of $\phi$, and
$h$ is an arbitrary function of $g_{\Omega}$.

Throughout the rest of this work we will choose these functions as
\bea
 f(\phi)  &=&  - \half \phi^2 ,\\
 h(g_{\Omega})  &=&  m^2 + V({g_{\Omega}}) .
\eea
We do that, so that the equation of motion for the scalar field becomes
\be  \label{eq_phi_sph}
  \left( \cov_a \cov^a - m^2 - V({g_{\Omega}}) \right) \phi = 0,
\ee
where we interpret the function $V(g_{\Omega}(t,r))$ as a
(coordinate-dependent) potential. The parameter $m$ is set to zero in our
simulations. The function $g_{\Omega}(t,r)$ is
just the square of the areal radius, $R(t,r)$. Then, we can
write \eq{eq_phi_sph} in the form
\be  \label{eq_phi_sph_2}
  \left( \cov_a \cov^a - m^2 - \tilde{V}(R) \right) \phi = 0,
\ee
where $\tilde V$ is an arbitrary function of the areal radius.

In appendix \ref{axial} we summarize the results obtained in the case
of axial symmetry.

%%%%%%%%%%%%%%%%%%%%%%%%%%%%%%%%%%%%%%%%%%%%%%%%%%
\section{The Equations} \label{TheEquations}
%%%%%%%%%%%%%%%%%%%%%%%%%%%%%%%%%%%%%%%%%%%%%%%%%%
In this work we solve the non-vacuum Einstein equations for a dynamic
spherically symmetric space time, coupled to a real scalar field. The
scalar field satisfies a Klein-Gordon-like equation with the addition of
a potential, as explained in section~\ref{sphericalcase}.

The equations are decomposed using a Cauchy formulation, in which the
space-time is foliated by space-like surfaces. The particular
formulation used is the Einstein-Christoffel hyperbolic formulation
\cite{york_fixing}, where the equations are decomposed into a system
of first order hyperbolic ``evolution equations,''  plus a system
of (first order)  ``constraint equations.'' These equations can be
solved by giving initial data that satisfy the constraint equations on
a given surface of the foliation, and then integrating the evolution
equations in time. The constraint equations at later times are then
automatically satisfied \cite{wald} in the domain of dependence of
that surface.

The equations solved are the Einstein-Klein-Gordon equations, 
with the addition of a potential,
\bea
  &G_{ab} = 8\pi T_{ab}&, \\
  &\left( \cov_a \cov^a  -V \right) \phi = 0&,   \label{eq_phi_sph_3}
\eea
where the stress-energy tensor, $T_{ab}$, and the potential, $V$, are
 given according to  section \ref{sphericalcase}, as well as the
 condition that $\phi$ is independent of $(\theta,\phi)$. In
 equation~\eq{eq_phi_sph_3} we have set $m=0$, but this parameter can
 be incorporated in the definition of $V$.
 
We consider the line element and extrinsic curvature of a space time in
spherical symmetry in the form 
\bea
  ds^2 &=& -N^2 dt^2 + g_{rr}(dr+\beta dt)^2 + r^2 g_T d\Omega^2, \\
  K_{ij}dx^idx^j &=& K_{rr} dr^2 + r^2 K_T d\Omega^2, \label{Kij}
\eea
where  $\beta$ is the ($r$ component of the) shift vector, and $N$ is the lapse
function. 
In the Einstein-Christoffel formulation, the shift and ``densitized
lapse'' function,  $\alpha\equiv N/\sqrt{g}$, are arbitrarily specified
and  kept fixed during the evolution. We denote by $g$ the determinant of
the three-metric.

In spherical symmetry, this system reduces to nine first order
evolution equations, and four first order constraint equations,
the later containing only spatial derivatives.

The variables evolved are: the metric components, $g_{rr}$ and
$g_T$; the scalar field, $\phi$; and other variables used to
convert the equations from second to first order. They are: the
extrinsic curvature components, $K_{rr}$ and $K_T$ (defined
in  eqn.\eq{Kij}); variables $\{\Psi, \Pi \}$ constructed with first-derivatives of $\phi$,
\bea
 \Psi &=& \d_r\phi, \\
 \Pi  &=& \frac{1}{N}\left(\beta \;\d_r\phi-\d_t\phi\right);
\eea
and the variables $\{f_{rrr},f_{rT}\}$ containing first spatial derivatives
of the metric,
\bea
 f_{rrr} &=& \frac{\d_r g_{rr}}{2} + \frac{4 g_{rr} f_{rT}}{g_T}, \\
 f_{rT}  &=& \frac{\d_r g_{T}}{2} + \frac{g_T}{r}.
\eea

The complete expressions of these equations are shown in detail 
in appendix~\ref{app_eq}. Their derivation, and the notation used, is
based on~\cite{Kidder} and~\cite{cpbc}, with the
addition of terms containing the potential.

%%%%%%%%%%%%%%%%%%%%%%%%%%%%%%%%%%%%%%%%%%%%%%%%%%
\section{Numerical Implementation} \label{numerics}
%%%%%%%%%%%%%%%%%%%%%%%%%%%%%%%%%%%%%%%%%%%%%%%%%%

%%%%%%%%%%%%%%%%%%%%%%%%%%%%%%%%%%%%%%%%%%%%%%%%%%
\subsection{Initial Data} \label{InitialData}
%%%%%%%%%%%%%%%%%%%%%%%%%%%%%%%%%%%%%%%%%%%%%%%%%%
Consistent initial data must satisfy equations \eq{C}-\eq{Cm}.
These equations determine some variables in terms of others
judiciously chosen. In this work, we exploit this freedom
to describe a black hole centered at $r=0$
by specifying $\{V,\phi,g_{rr},K_{rr}\}$ from the known Schwarzschild solution
and solving for $g_T$ and $K_T$.

Before describing the details of our implementation, we discuss
how the potential and scalar field are chosen.
We adopt a potential $V$ with two free parameters $\{A,r_0\}$ to 
regulate the depth and location of the ``well'' where the
scalar field is to be confined (see figure~\ref{potentials}). A simple
expression for $V$ suffices for this task, and we adopt
\be   \label{V}
V(R) = A\left(1-e^{-\left(R-r_0\right)^2}\right),
\ee
with the areal radius $R$ given by $R=r\sqrt{g_T}$.
The parameters in
this expression were set to $A=30/M^2$ and $r_0=6M$, where $M$ is the
initial mass of the black hole.
Notice that
during the evolution $R=R(t,r)$, thus, in these coordinates, 
the shape (and position) of the potential well can change in time. 
We will return to this point later.

The scalar field $\phi$ is defined following either one of two different
strategies. One is designed to conform to time-harmonic situations
in weakly-gravitating cases and the other simply prescribing a
sufficiently smooth
profile. The latter choice allows us to investigate the spacetime's response
to fields not designed to conform to a time-harmonic dependence. 

%%%%%%%%%%%%%%%%%%%%%%%%%%%%%%%%%%%%%%%%%%%%%%%%%%
\subsubsection{Time-harmonic scalar field} \label{statid}
%%%%%%%%%%%%%%%%%%%%%%%%%%%%%%%%%%%%%%%%%%%%%%%%%%
To prescribe a scalar field which will give rise to a spacetime with harmonic time-dependence,
we begin by considering the  limiting case when the scalar field's amplitude
is negligible; there the metric should be described 
by the Schwarzschild's solution.
Now, considering the scalar field as existing over this fixed background spacetime,
a Schr{\" o}dinger-like eigenvalue equation can be obtained to 
determine time-harmonic states as discussed
 below. 

The Schwarzschild metric in Eddington-Finkelstein coordinates is:
\bea
 ds^2 &=&- \left(1-\frac{2M}{r}\right) dt^2
         + \left(1+\frac{2M}{r}\right) dr^2 +\nonumber \\
       &&+ \frac{4M}{r} dtdr + r^2 d\Omega^2 . \label{Sch}
\eea
We use this metric to evaluate the equation of motion for $\phi$,
equation~\eq{eq_phi_sph_3}. To solve this PDE we
use the following ansatz  that yields separation of
variables\footnote{Suggested by the fact that in Schwarzschild coordinates,
  ($\tilde{t}$, $\tilde{r}$), the ansatz $\phi =
  u(\tilde{r})\cos(\omega \tilde{t})$ yields separation of
  variables. The coordinates transformation being:
  $\tilde{t}=t-2M\ln\left(\frac{r-2M}{M}\right)$, $\tilde{r}=r$.},
\be \label{separation}
 \phi(t,r) = u(r) \cos\left( \omega\left[t-2M\ln\left(\frac{r-2M}{M}\right)\right] \right).
\ee
The equation for $u(r)$ results
\be \label{Lu}
 \mathcal{L}\; u(r) = \left[\omega^2 - 
 \left(1-\frac{2M}{r}\right)V(r) \right]u(r),
\ee
where the second order operator $\mathcal{L}$ is given by
\bea
 \mathcal{L}&=&-\left(1-\frac{2M}{r}\right)^2
  \frac{\d^2}{\d r^2} \nonumber \\
    && -\frac{2}{r} \left(1-\frac{M}{r}\right) \left(1-\frac{2M}{r}\right) \frac{\d}{\d r}
\eea
Equation \eq{Lu} is integrated to obtain $u(r)$. Then, from its
  definition, equation \eq{separation},  
$\phi(t,r)$ is calculated. Finally, from $\phi(t,r)$ we obtain $\Pi(t,r)$ and
 $\Phi(t,r)$  evaluating these functions at $t=0$ and adopting them as initial data.

Equation \eq{Lu} can be straightforwardly integrated to obtain both the eigenvalue
and eigenfunction through a standard shooting algorithm.
To this end, we transform the second-order equation to a system of two
first order equations for $u(r)$ and $u'(r) \equiv d u /dr$ augmented
with a third equation ($(\omega^2)'=0$) 
to simplify the implementation (see \cite{nrf} for the details).

The system of equations is then
integrated outwards from $r_{L}\equiv 4M$ on one hand, and also inwards
from $r_{R}\equiv 8M$.
The obtained solutions are matched at an intermediate point, in our case at $r_0$ (the center
of the potential well), with the conditions that both the solutions
and derivatives are continuous.
The initial guesses for the boundary conditions are then
varied until a satisfactory match is obtained. The code used to
implement the shooting algorithm is
the one described in \cite{nrf}, except that the ODE integrator is
replaced for LSODE (Livermore Solver for ODEs) \cite{lsode}.
The boundary conditions, consistent with the physical scenario in mind are determined 
as follows.

\begin{figure}[!tbh]
  \includegraphics[angle=0,width=\columnwidth,height=!,clip]{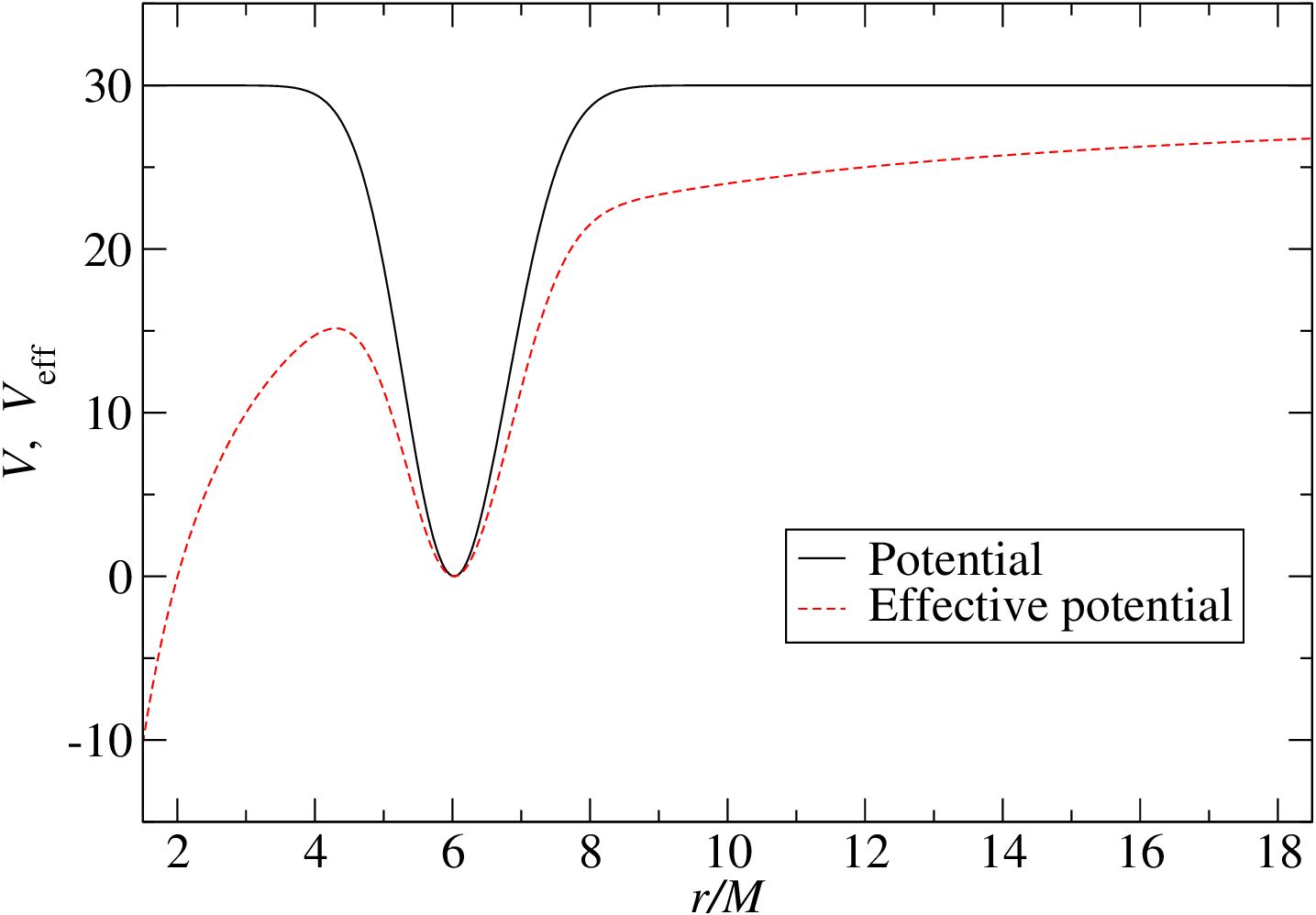}
  \caption{Potential and effective potential, as defined in~\eq{V}
  and~\eq{Veff}, respectively. As mentioned in the text, the
  potentials are, in general, functions of $R\equiv r\sqrt{g_T}$. The
  potentials showed in this figure are those used to find the
  time harmonic states $u(r)$, where the Schwarzschild metric is used,
  hence $R=r$.
  \label{potentials} }
\end{figure}

We have a system of three first order ODEs, thus
three boundary conditions need be specified. 
Natural conditions for our purposes result from
requiring the fields fall sufficiently rapid
at the boundaries. We thus impose a relationship
between $u$ and its derivative at each boundary, 
of the form $u'=k u$. The coefficient $k$ at each boundary
can be found through a WKB-type approach. 
To do so, we first consider the variable change $u(r) \equiv F(r) \tilde u(r)$ and
fix $F(r)=[r(r-2M)]^{-\half}$ so as to remove the first order derivative in equation~\eq{Lu}.
The resulting equation is
\be
-f(r)\tilde{u}''(r)+ V_{\rm eff}(r) \tilde{u}(r)= \omega^2\tilde{u}(r)
\ee
with $f(r)$ and $V_{\rm eff}(r)$, 
\begin{eqnarray}
f(r) &=&\left(1-\frac{2M}{r}\right)^2 ,\\
V_{\rm eff}(r) &=& \left(1-\frac{2M}{r} \right) V(r) -
\frac{M^2}{r^4}, \label{Veff}
\end{eqnarray}
and we interpret $V_{\rm eff}$ as an effective potential
(which is shown in figure~\ref{potentials}).
Next, we freeze the coefficients  $f(r)$ and $V_{\rm eff}$ on a small neighborhood 
of each boundary point and consider solutions of the form $\exp(\pm k r)$, with $k^2=(V_{\rm
  eff}-\omega^2)/f$. The following conditions at the boundaries are then determined by
\bea
&&\tilde{u}(r) \propto  e^{+k_1 r}
                 \quad \mathrm{at} \quad r=r_{L} ,\\
&&\tilde{u}(r) \propto e^{-k_2 r}
                 \quad \mathrm{at} \quad r=r_{R},
\eea
where
\bea
&&k_1=\sqrt{\frac{V_{\rm eff}(r_{L})-\omega^2}{f(r_{L})}} ,\\
&&k_2=\sqrt{\frac{V_{\rm eff}(r_{R})-\omega^2}{f(r_{R})}} .
\eea
As illustrated later, these conditions indeed ensure the solutions decay
rapidly outside of the potential well (for a bounded range of values of $\omega^2$).
Notice that since the equations are homogeneous there remains a freedom on 
the amplitude of the fields at the boundaries. We fix this freedom by setting 
$u=1$ at $r_{L}$ and adopting as the varying parameter for the
shooting method the value of $u$ at $r_{R}$.

Once obtained $\phi(r)$ in $[r_{L}, r_{R}]$ using equation
\eq{separation}, we set $\phi(r)=0$ outside this region. For
the amplitude of $\phi$ used in this work in the case of time-harmonic
initial data, the values of $\phi$ and its derivative at $r_{L}$ and
$r_{R}$ are small enough to ensure that this matching is sufficiently smooth, as is
corroborated when evolving these initial data.

\subsubsection{Smooth profile} \label{pulse}

The other approach employed in this work is to adopt a simple
expression for the scalar field. In particular we adopt a
``pulse'' of compact support of the form
\be   \label{pulsedef}
 \phi(r) = \left\{
          \begin{array}{cc}
          c (r-r_1)^4 (r-r_2)^4 & r_1\le r \le r_2 \\
	  0                     & \mathrm{elsewhere}
          \end{array}
	  \right. ,
\ee
where the
values $r_1$ and $r_2$ control the width of the pulse and were 
chosen so that it is centered with the potential (at $r=6M$):
$r_1=5M$, $r_2=7M$. After specifying $r_1$ and $r_2$, the coefficient
$c$ is chosen so that the scalar field has a given mass. 
This initial data is used to compare with the previous approach in
regimes where the fixed-background approximation is justified and to study
the spacetime's behavior in non-linear cases.

\subsubsection*{Remaining data}
Having specified both the potential and the scalar field, consistent initial
data is determined by integrating the constraint
equations in the following manner. First, the functions $g_{rr}$, $K_{rr}$, $\alpha$, and
$\beta$ are set equal to those  read-off from the Schwarzschild solution in
Eddington-Finkelstein coordinates. Adopting these coordinates gives the freedom to
place the inner boundary inside the black hole.
We found it convenient to rewrite the constraint equations  in the form:
\bea
 \d_r g_T &=& d_T ,\\
 \d_r  d_T &=& f_1(g_T,d_T,K_T;F_i) ,\\
 \d_r K_T &=& f_2(g_T,d_T,K_T;F_i) ,
\eea
where $F_i$ represents all the functions that are specified a priori
(including $\phi$).
These equations are integrated outwards from the inner boundary using the step
adaptive integrator LSODE,
using as boundary data ($g_T$, $d_T$, and $K_T$ at $r=r_{\rm min}$) the
values read-off from the Schwarzschild solution.

%%%%%%%%%%%%%%%%%%%%%%%%%%%%%%%%%%%%%%%%%%%%%%%%%%
\subsection{Evolution}
%%%%%%%%%%%%%%%%%%%%%%%%%%%%%%%%%%%%%%%%%%%%%%%%%%
We discretize the equations with a scheme formulated to take advantage 
of numerical techniques which guarantee stability of generic linear 
first order hyperbolic systems. In this work we
adopt: (i) second order accuracy by implementing second-order
derivative operators satisfying summation by parts~\cite{KS1,KS2,strand,SBP0,SBP1}; 
(ii) a third-order 
Runge-Kutta operator for the time integration through the method of
lines~\cite{tadmor};
(iii) a Kreiss-Oliger~\cite{KO} style dissipative algorithm to control 
the high frequency
modes of the solution~\cite{gustaffsonkreissoliger,SBP1,SBP2} and 
(iv) maximally dissipative boundary conditions setting
all incoming modes to zero~\cite{olsson,gustaffsonkreissoliger}. 

We employ a uniform grid to cover the region $r\in [r_{\rm min},r_{\rm max}]$ with
$N$ equi-spaced points. The grid-spacing between points is $\Delta
r = (r_{\rm max} - r_{\rm min}) / (N-1)$. The time step $\Delta t$ is defined in terms 
of $\Delta r$ as $\Delta t= cfl\; \Delta r$ and $cfl=0.25$ is chosen so that
the CFL condition \cite{thomas} is satisfied. In what follows,
sub-indices denote particular points of a slice, and super-indices
distinguish each slice. 

The inner boundary, $r=r_{\rm min}$, is set inside the black hole initially,
and monitored during the evolution to ensure that it remains
inside and constitutes and outflow boundary of the computational domain. 
Then, there is no need to prescribe boundary conditions there. At
the outer boundary, $r=r_{\rm max}$ maximally dissipative boundary conditions are adopted.
In our present case we take the simplest form of these conditions and
set the incoming modes to zero. 
The characteristic structure for the system of equations is detailed
in appendix \ref{characteristic}.

The code have been tested to ensure that the numerical solutions
obtained converge to the corresponding solutions of the Einstein
equations. In appendix~\ref{tests} we show the convergence test for
the Hamiltonian constraint.

%%%%%%%%%%%%%%%%%%%%%%%%%%%%%%%%%%%%%%%%%%%%%%%%%%
\section{Analysis and Results} \label{results}
%%%%%%%%%%%%%%%%%%%%%%%%%%%%%%%%%%%%%%%%%%%%%%%%%%
In the simulations performed in this work we set the initial mass of the black hole to $M=1$
(in geometrized units). The domain of integration was chosen so that the
region of interest is unaffected by the conditions adopted at
the right boundary. This corresponds to $r_{\rm min}=1M$ and $r_{\rm max}=221M$. The maximum
resolution used was $\Delta r=0.01M$ (22000 grid points).

In the two approaches we use to obtain initial data, we have the
freedom of adjusting the amplitude of the scalar field, which in turn
determines its mass. We set initial data where the mass of the scalar
field is $m_{\rm sf}=0.01M$ in the time-harmonic case, while for the non-time-harmonic
cases we set $m_{\rm sf}$ equal to $0.01M$, $\kappa\,0.1M$, ($M$ being the
initial mass of the black hole and $\kappa = 1...5$). To calculate
the mass we use the Misner-Sharp formula \cite{MTW},
\be  \label{MSdef}
 M_{\rm MS}(r) = \frac{r\sqrt{g_T}}{2}
  \left[1+\frac{r^2}{g_T}\left(K_T^2-\frac{f_{rT}^2}{g_{rr}}\right)\right],
\ee
which measures the total mass inside a spherical surface
labeled by coordinate $r$. In our initial data the mass of the black
hole, $M$, is preset, so we can calculate $m_{\rm sf}$ by subtracting $M$ from
the total mass of the space-time,
\be
m_{\rm sf} = M_{\rm MS}(R) - M ,
\ee
where $R$  labels a sphere containing the scalar field, which is
localized initially. (See figure~\ref{mass001st}).

During the evolution we employ this formula, replacing
$M$ for $M_{\rm MS}$ at the horizon\footnote{The position of
 the apparent horizon is given by the outermost trapped surface.}. 

In our analysis we also evaluate the Kretschmann invariant $I\equiv
R_{abcd}R^{abcd}$, where $R_{abcd}$ is the Riemann tensor. This
quantity provides a gauge-invariant answer that can be compared with
its value in known spacetimes. For
a Schwarzschild space-time, $I$ is given by 
\be \label{KI}
I_{\rm Sch} = \frac{48 {(M_{\rm MS})}^2}{R^6},
\ee
where, in Schwarzschild coordinates, $M_{\rm MS}=M$ and $R=r$. We evaluate
the quotient $I/I_{\rm Sch}$ using \eq{KI} with
$R=r\sqrt{g_T}$ and $M_{\rm MS}$ defined in \eq{MSdef}.

%%%%%%%%%%%%%%%%%%%%%%%%%%%%%%%%%%%%%%%%%%%%%%%%%%
\subsection{Initial Data}
\begin{figure}[!tbh]%[!tbh]
  \includegraphics[angle=0,width=0.9\columnwidth,height=!,clip]{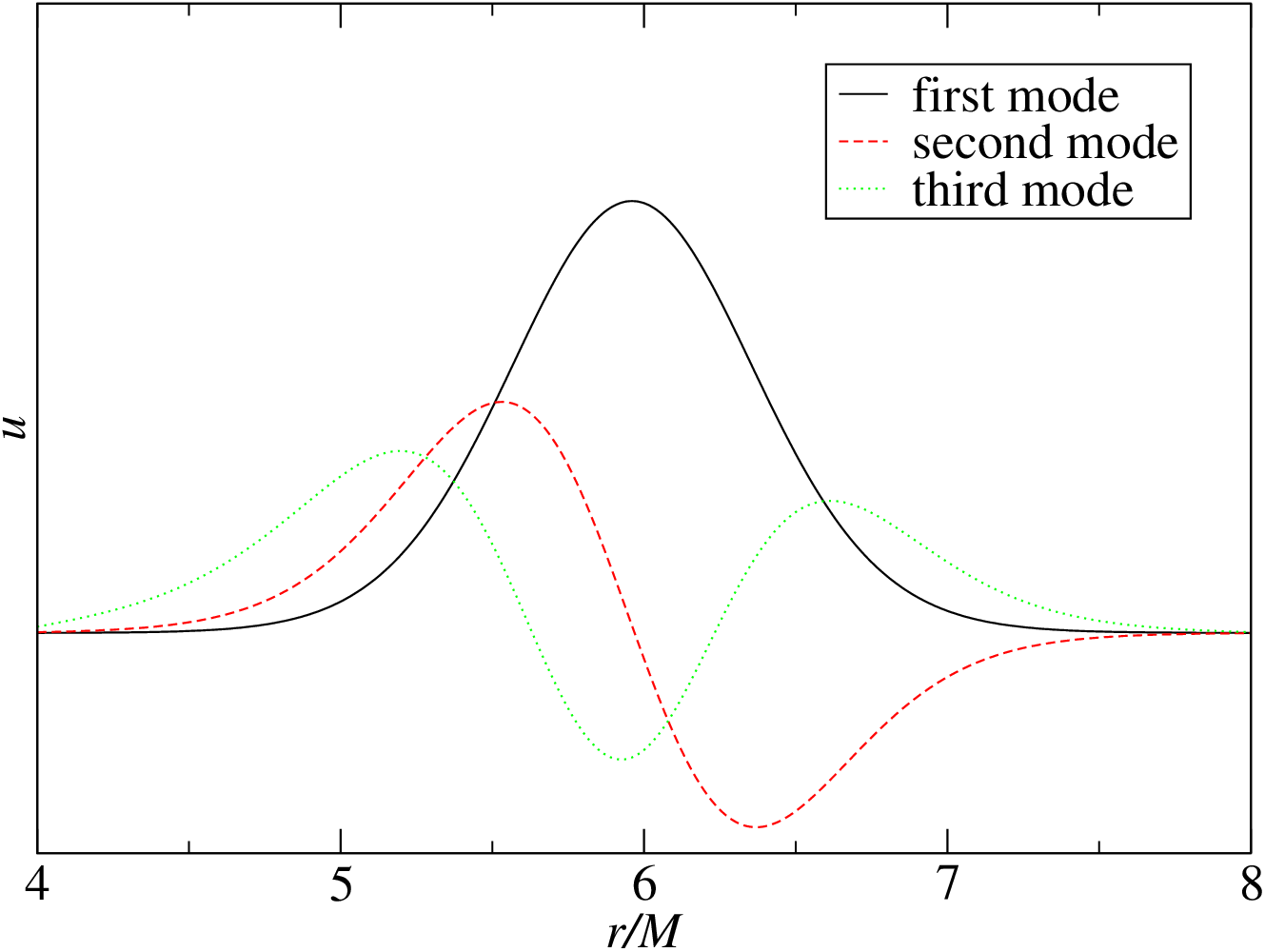}
  \caption{First time-harmonic states of $u(r)$. \label{u_modes} }
\end{figure}

\begin{figure}[!tbh]%[!tbh]
  \includegraphics[angle=0,width=\columnwidth,height=!,clip]{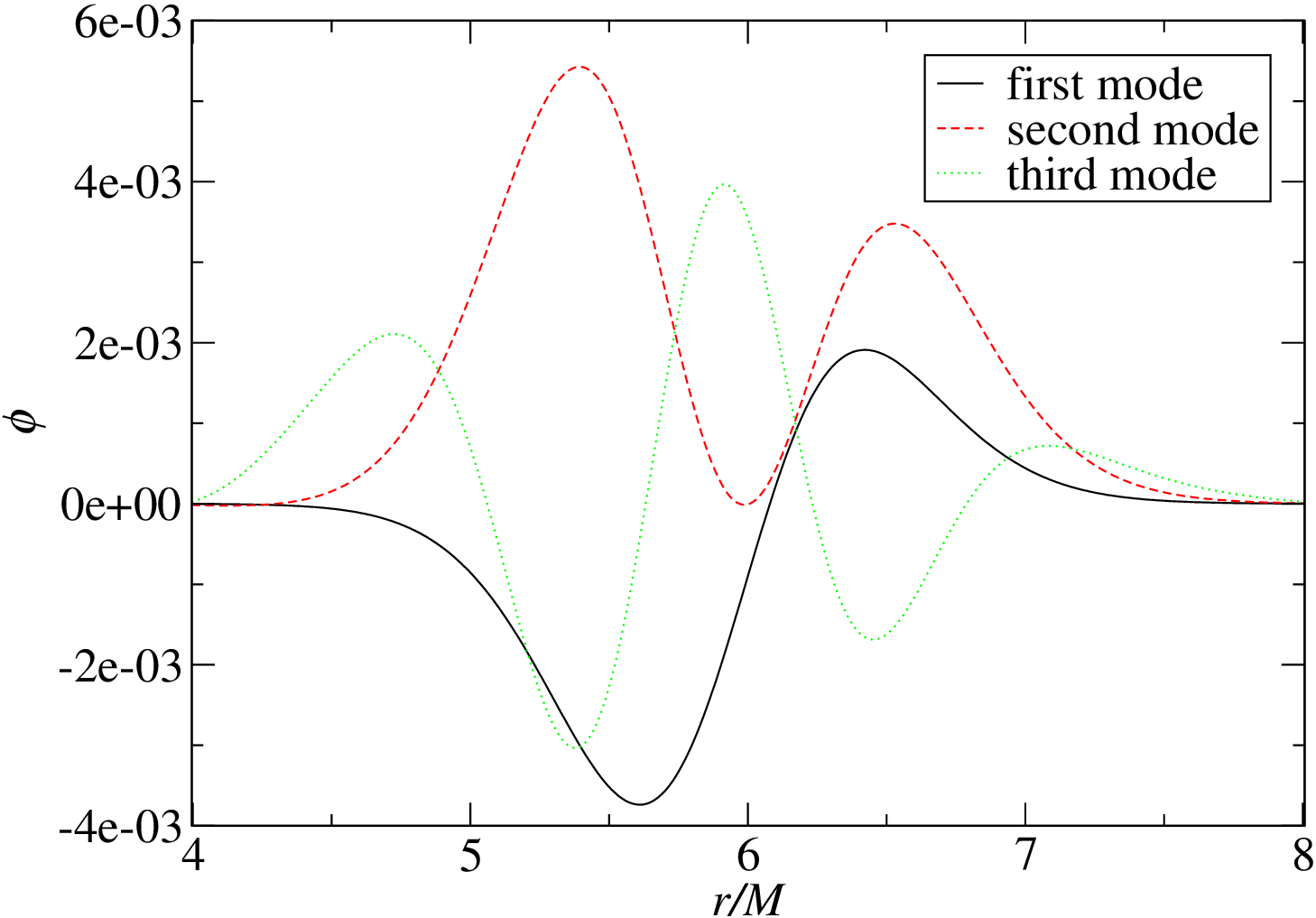}
  \caption{Scalar field at $t=0$ obtained from the first time-harmonic
  states of $u(r)$, using equation~\eq{separation}.  \label{phi_modes} } 
\end{figure}

%%%%%%%%%%%%%%%%%%%%%%%%%%%%%%%%%%%%%%%%%%%%%%%%%%
As explained in section~\ref{statid}, we first find time-harmonic states for
the scalar field on a Schwarzschild space-time. By varying the
initial guess for the frequency in the shooting integration we obtain
different modes. We show the first modes in figures~\ref{u_modes} and~\ref{phi_modes}.
However, for this work we used only the first mode which will be referred to as
 ``the time-harmonic state'', unless otherwise specified.
These modes have been re-scaled so that they can be normalized (in analogy with quantum
  mechanics) so that $\int{r^2 |u(r)|^2 dr=1}$. There is no physical
  justification for choosing that 
  particular normalization, but it is helpful when comparing different
  eigenstates, which otherwise would have greatly different
  amplitudes.


The other approach used to define the initial data corresponds to
the ``pulse'' described in section~\ref{pulse}.
In the linear regime we employ both types of initial
data, with a scalar field's initial mass $m_{\rm sf}=0.01M$. 
In the non-linear regime we adopt only the non-time-harmonic initial data with
masses $m_{\rm sf}$ ranging from $0.1M$ to $0.5M$. 

%%%%%%%%%%%%%%%%%%%%%%%%%%%%%%%%%%%%%%%%%%%%%%%%%%
\subsection{Evolution}
%%%%%%%%%%%%%%%%%%%%%%%%%%%%%%%%%%%%%%%%%%%%%%%%%%
We study the evolution of the prescribed data. We begin by
considering  first the linear regime, adopting scalar field configurations
with initial mass of  1\% of that of the black hole.  After confirming that
the time-harmonic configuration behaves as expected, we confirm that the ``pulse'' configuration
evolves towards a time-harmonic regime.  Then, we study cases in the 
non-linear regime, with initial
scalar field masses ranging from 10\% to 50\% of that of the black hole. In all
cases we evolve until $t=200M$.

\subsubsection*{Linear case}
\begin{figure*}[!tbh]
\subfigure[]{
  \includegraphics[angle=0,width=\columnwidth,height=!,clip]{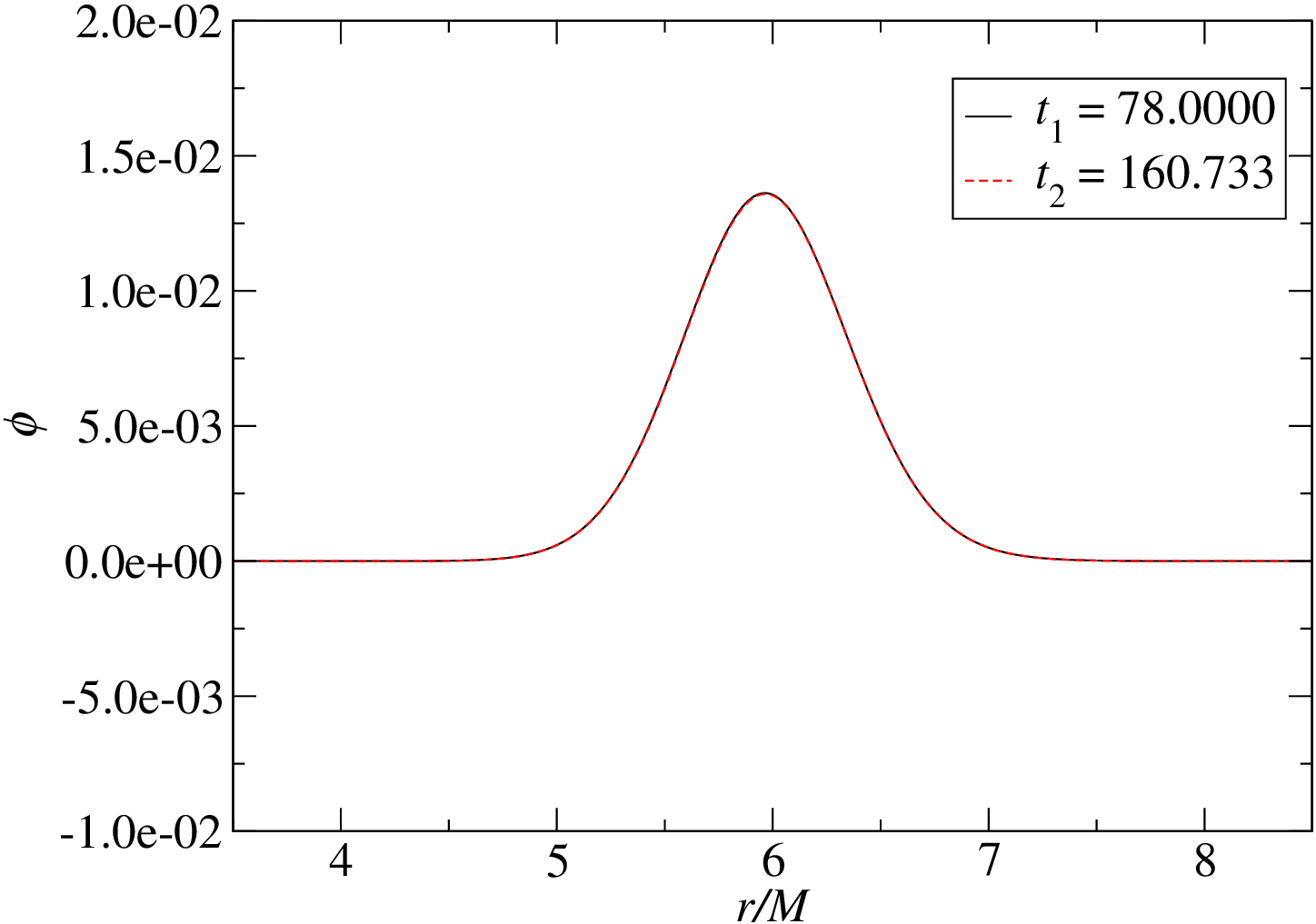}
\label{st001_a} }
\subfigure[]{
  \includegraphics[angle=0,width=\columnwidth,height=!,clip]{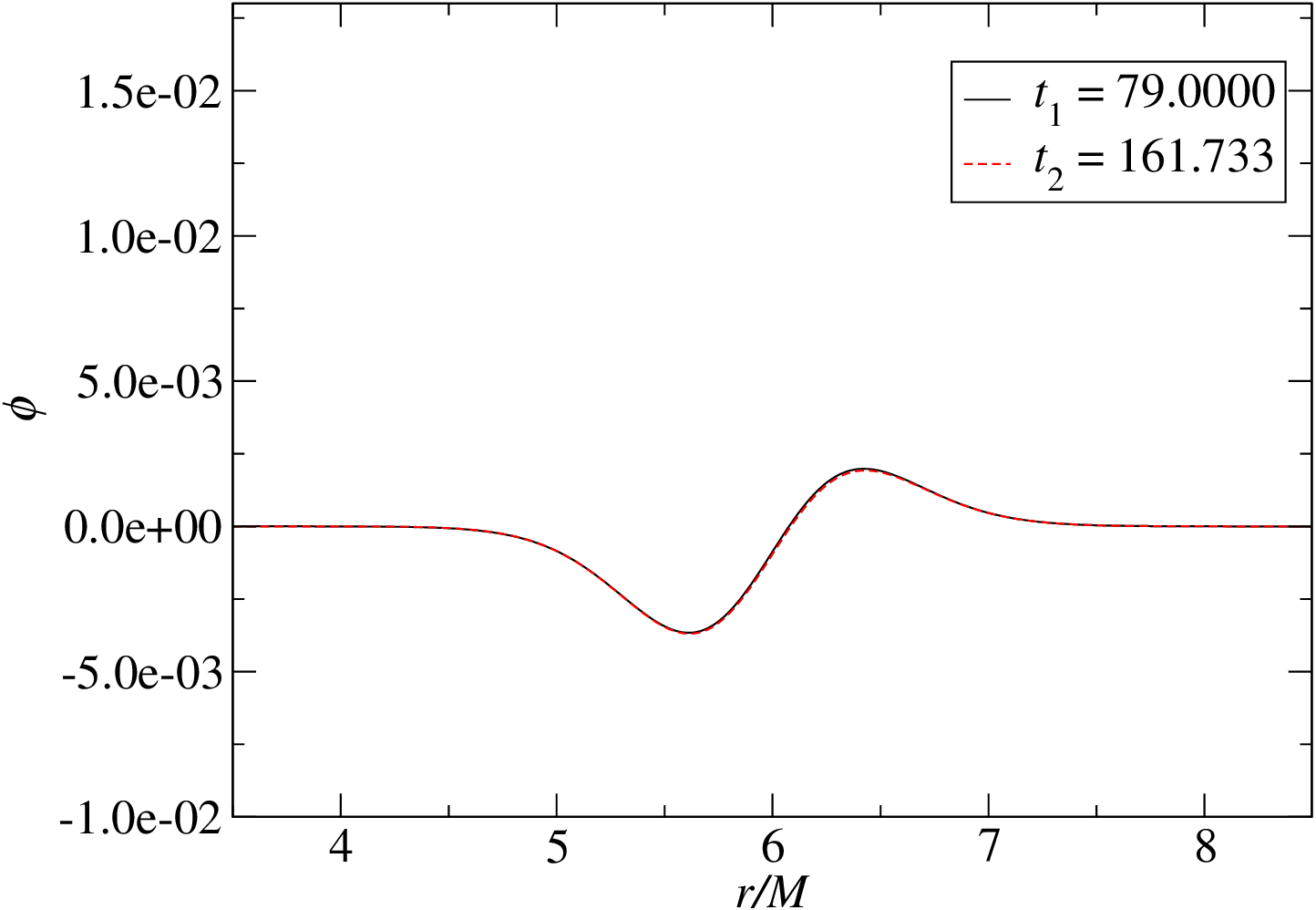}
\label{st001_b} }
\caption{The scalar field at different times is compared to check if the
  evolution remains described by a time-harmonic dependence. 
  Case with time-harmonic initial data. Initial mass of the 
  scalar field $m_{\rm sf}=0.01M$. Figure~\ref{st001_a} shows the scalar
  field when it reaches a maximum, while figure~\ref{st001_b} shows it at
  about a quarter of a period later. In both cases, the profile shown in continuous
  line is separated 22 periods from the one in dashed line.
\label{st001} }
\end{figure*}

\begin{figure*}[!tbh]
\subfigure[]{
  \includegraphics[angle=0,width=\columnwidth,height=!,clip]{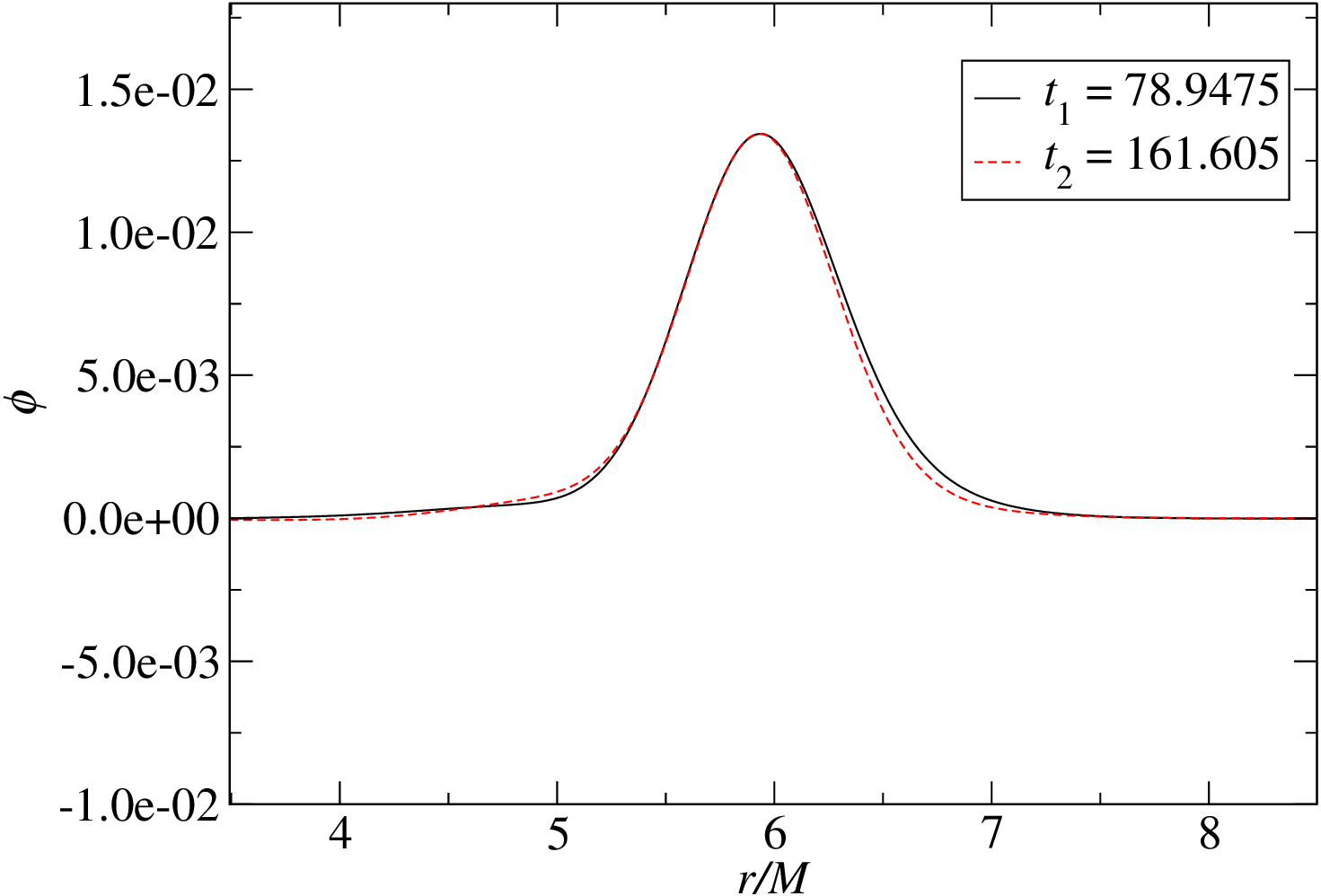}
\label{ns001_a} }
\subfigure[]{
  \includegraphics[angle=0,width=\columnwidth,height=!,clip]{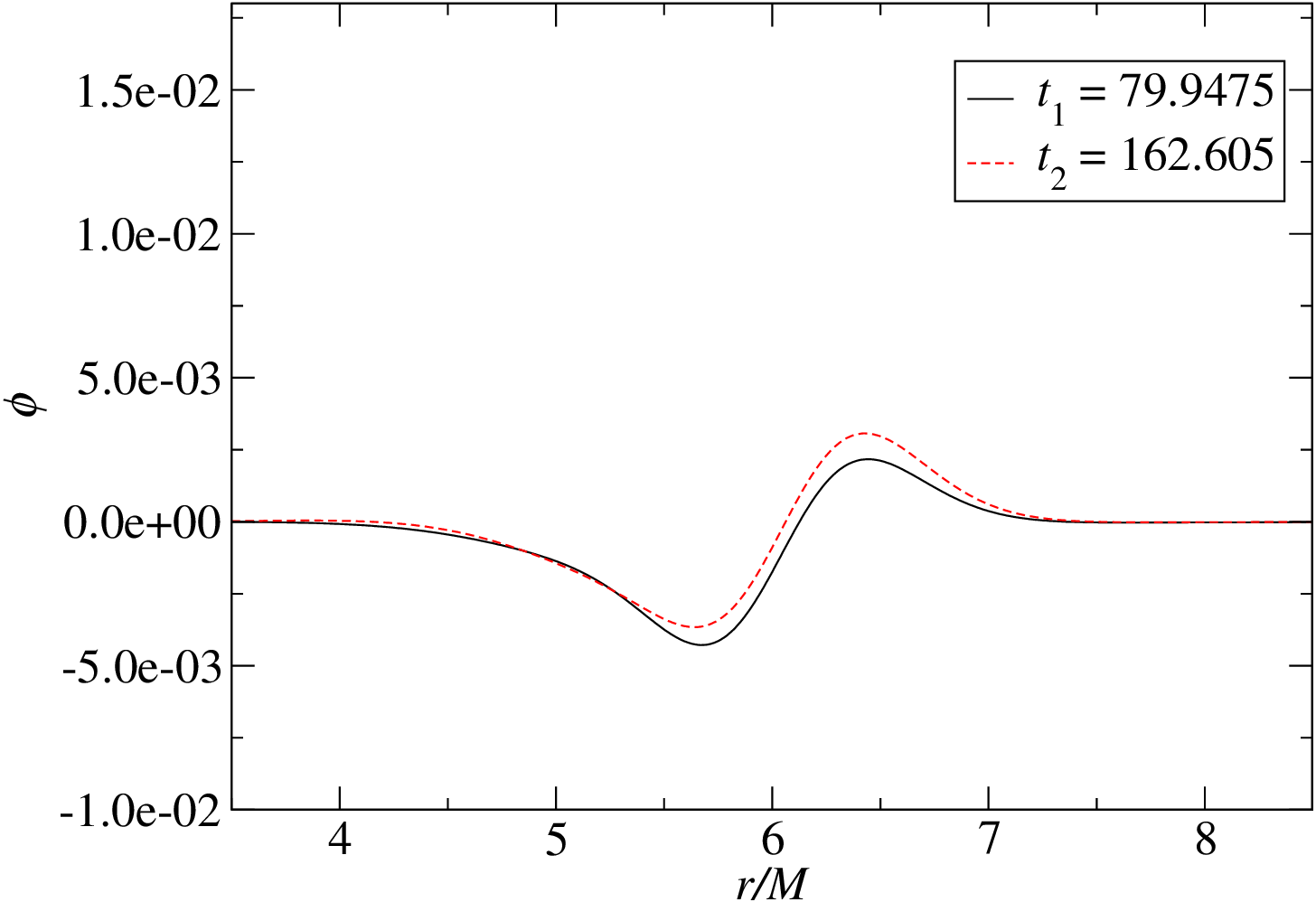}
\label{ns001_b} }
\caption{Here we show the same comparison of profiles as in figure~\ref{st001},
  this time for the case with non-time-harmonic initial data. The
  separation between the profiles compared is also 22 periods. The
initial  mass of the scalar filed is $m_{\rm sf}=0.01M$.
\label{ns001} }
\end{figure*}

\begin{figure*}[!tbh]
\subfigure[$\left|\phi(t_1)-\phi(t_2)\right|$ for the case with time-harmonic
  initial data (see figure~\ref{st001_a})]{
  \includegraphics[angle=0,width=\columnwidth,height=!,clip]{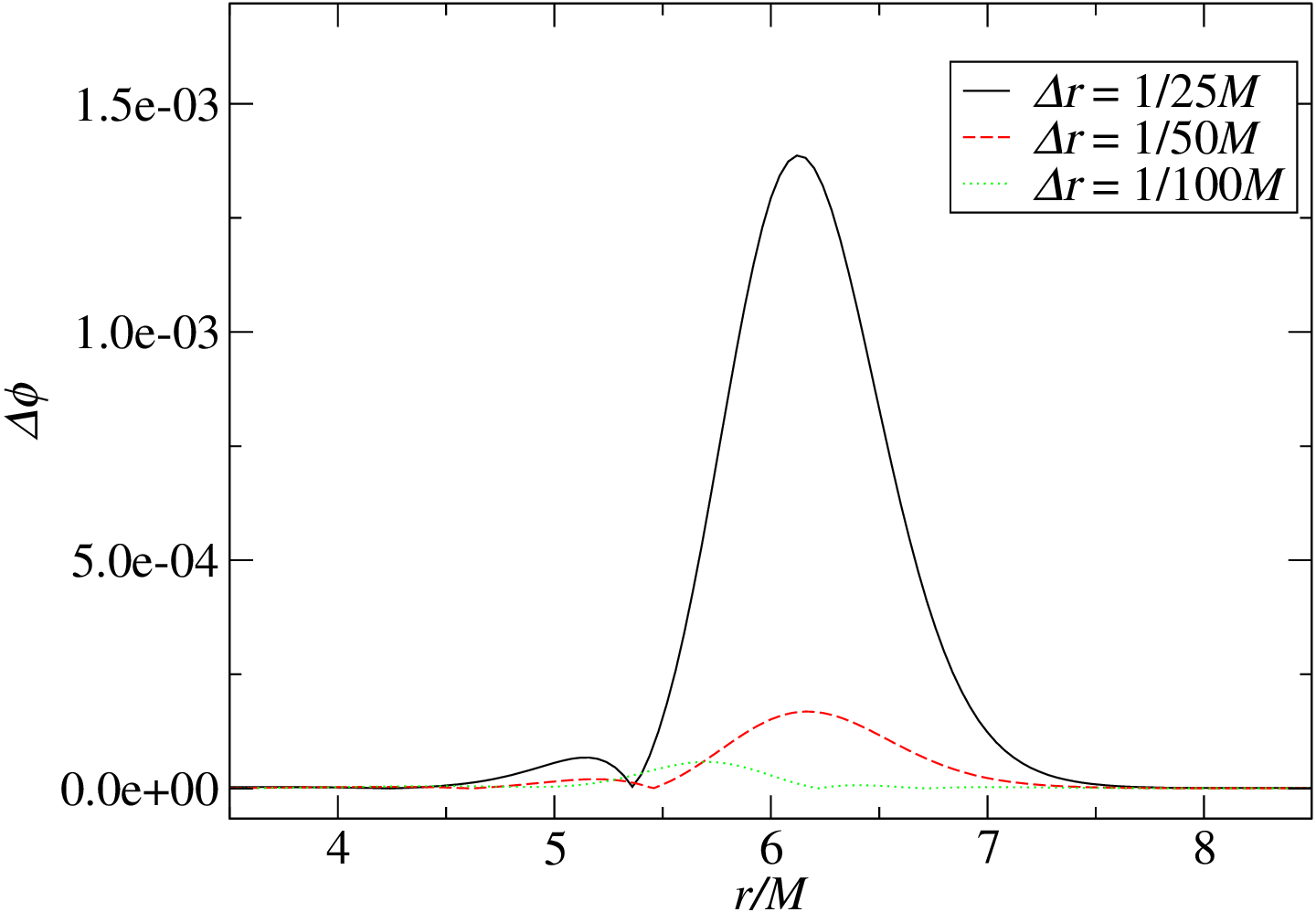}
\label{st_diff} }
\subfigure[$\left|\phi(t_1)-\phi(t_2)\right|$ for the case with non-time-harmonic
  initial data (see figure~\ref{ns001_a})]{
  \includegraphics[angle=0,width=\columnwidth,height=!,clip]{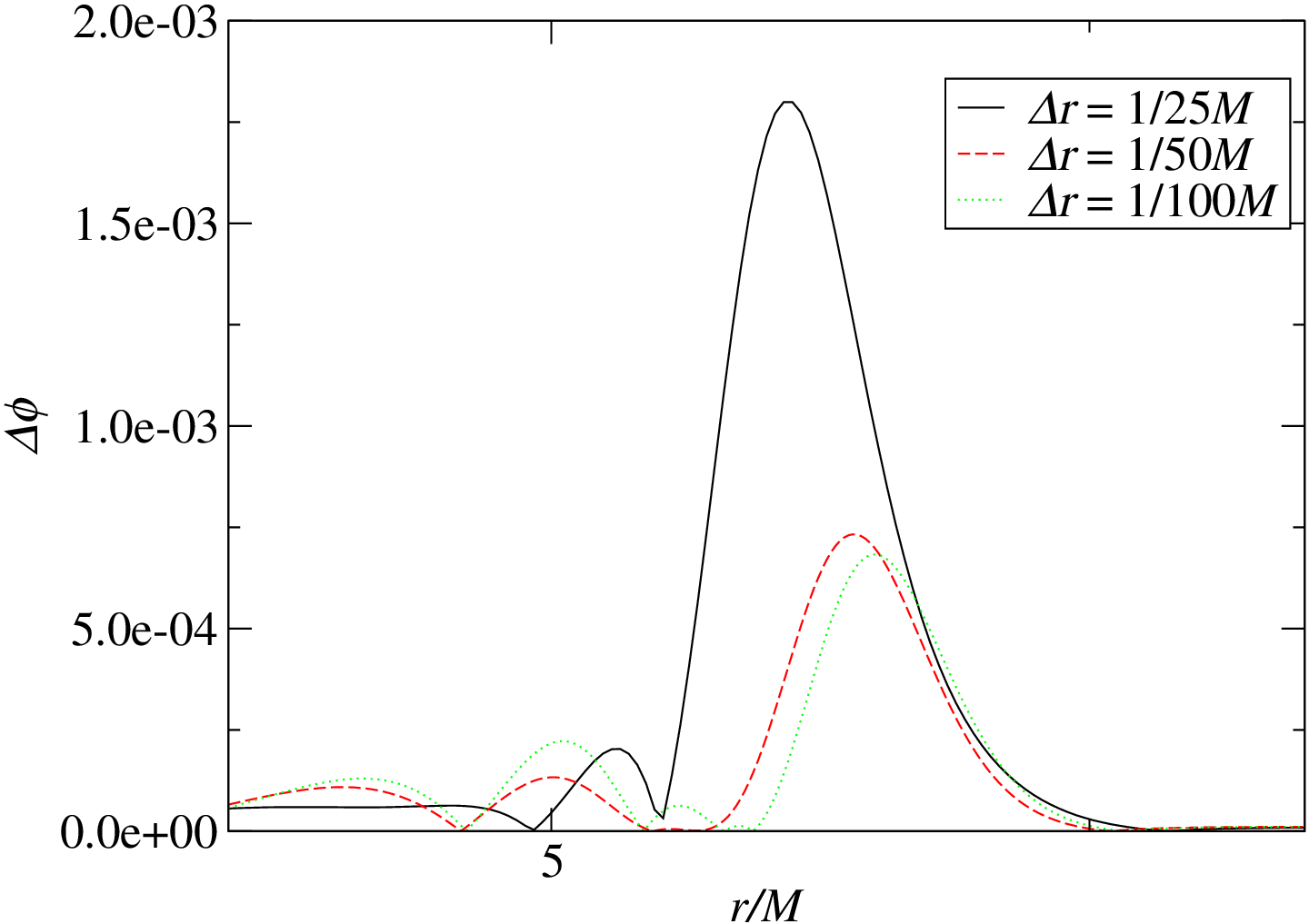}
\label{ns_diff} }
\caption{Absolute value of the difference between the scalar field at
different times: $\left|\phi(t_1)-\phi(t_2)\right|$, where
$t_2-t_1=22$~periods. Figure~\ref{st_diff} shows the 
difference between the profiles shown in figure~\ref{st001_a}, while
figure~\ref{ns_diff} shows the difference between those in
figure~\ref{ns001_a}. In each case, we show these differences for three resolutions.
 \label{diff} }
\end{figure*}

The time-harmonic initial data constructed essentially remains unchanged through
the evolution while the non-time-harmonic data evolves towards a time-harmonic state. 
Figures~\ref{st001} and~\ref{ns001} illustrate $\phi(r)$ at different times for
the maximum resolution employed ($\Delta r = M/100$).
Figure~\ref{st001} corresponds to the time-harmonic initial data,
and figure~\ref{ns001} to non-time-harmonic initial data.
In both cases we sampled along two different periods at $t\approx80M$; and
then at $t\approx160M$. The corresponding
pairs, are then plot together illustrating how after 22 periods apart the solutions
are essentially the same.
%
%%% FIG %%% Profiles. Stationary case (msf=0.01)
%
%%% FIG %%% Profiles. Non-stationary case (msf=0.01)
This is further illustrated in figure~\ref{diff} where we show the difference 
between each of these pairs for three different resolutions. 
%
%%% FIG %%% Profiles difference for three resolutions. Stationary case.

\begin{figure}[!tbh]
  \includegraphics[angle=0,width=\columnwidth,height=!,clip]{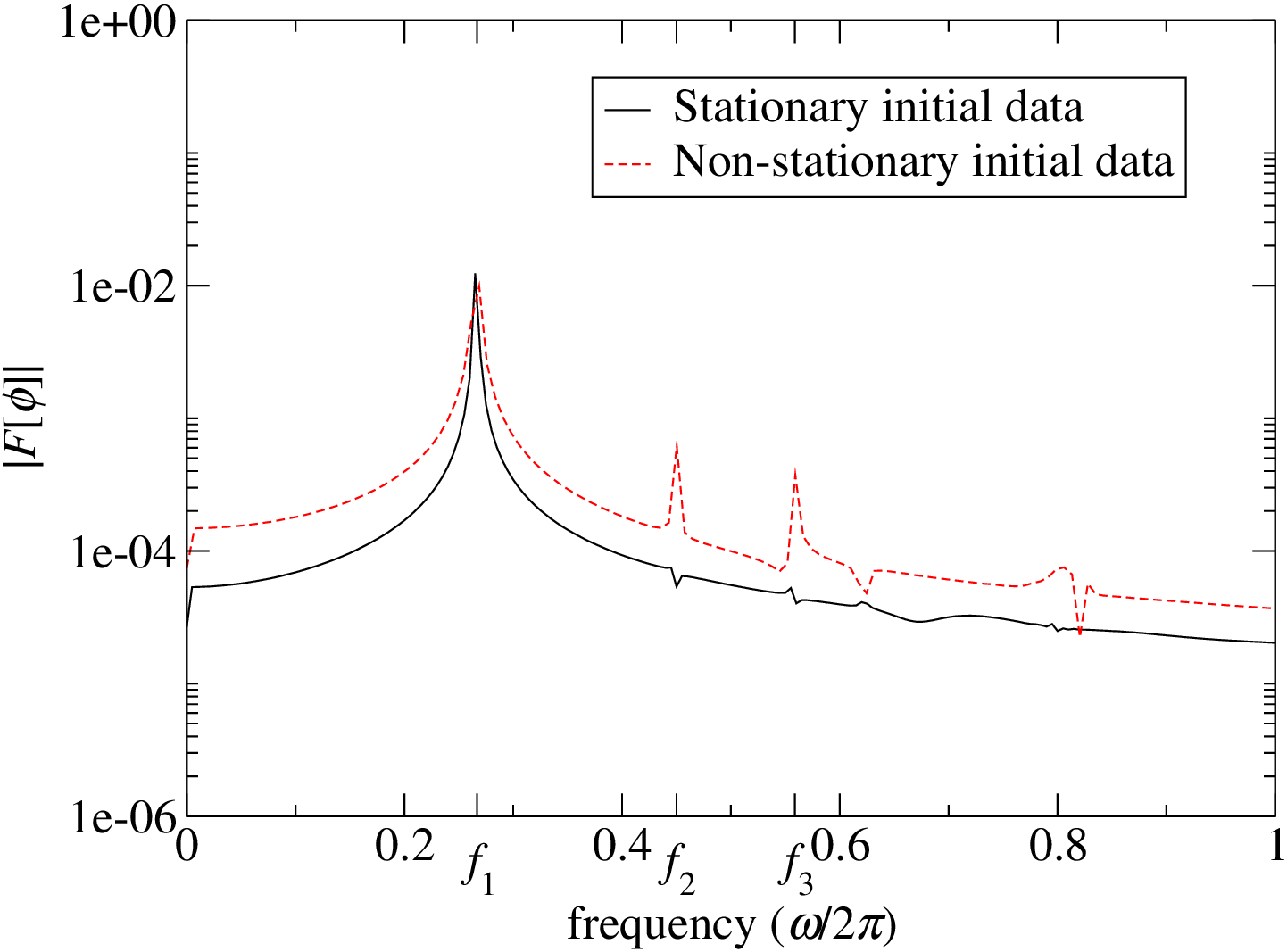}
  \caption{Absolute value of the discrete Fourier transform in time of $\int \! \phi
  \;dr$. The continuous line corresponds to the time-harmonic initial
  data, while the dashed line corresponds to the non-time-harmonic
  initial data. In the later case, the scalar field relaxes to
  a superposition of the first time-harmonic modes, whose frequencies are
  shown in the figure (labeled $f_n$). \label{DFT_001} }
\end{figure}

Finally, figure~\ref{DFT_001} displays the absolute value of the Fourier transform in time of
$\int \! \phi\;dr$, denoted $|F[\phi]|$. The scalar field is first integrated in space, then a discrete
Fourier transform in $t$ is calculated, where $t$ ranges from $0$ to
$200M$ in the case of time-harmonic initial data, and from $t_0=60M$ to
$200M$ in the non-time-harmonic case. In the plot we also indicate the frequencies
($f_n=\omega_n/2\pi$) obtained from the shooting integration when
calculating the time-harmonic states. The time $t_0$ is chosen after 
the initial transient behavior, indicated by a time-harmonic behavior
observed in $\phi$.
The initially non-time-harmonic scalar field relaxes to a
superposition of the first three time-harmonic modes, the first one
being the dominant one. We point out here that for this configuration,
the shooting method gives raise to three possible modes. It is thus
no surprising that the evolution gives rise to a solution described by these modes.
Deeper potentials give rise to more modes.

\begin{figure}[!tbh]
  \includegraphics[angle=00,width=\columnwidth,height=!,clip]{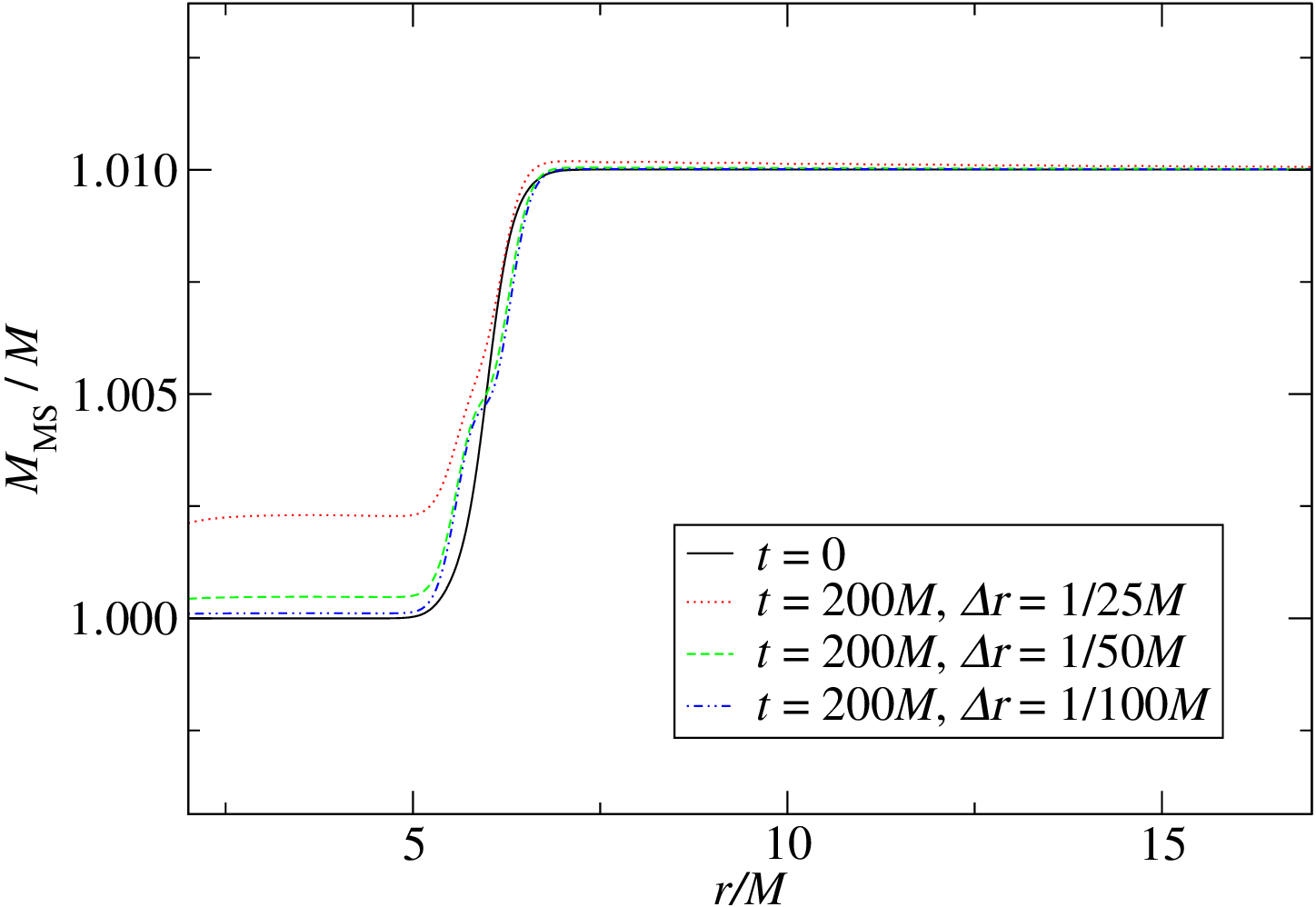}
  \caption{ Mass function at $t=0$; and at $t=200M$ for three
  resolutions. Stationary initial data. Initial $m_{\rm
  sf}=0.01M$. The continuous line shows the mass function at $t=0$,
  while the discontinuous lines show, for different resolutions, the
  mass function at $t=200M$. In
  this case the escape of mass into the black hole is negligible ($\Delta m_{\rm
  sf}=(1\pm3)\times 10^{-4}M)$. 
\label{mass001st}} 
\end{figure}

\begin{figure}
  \includegraphics[angle=0,width=\columnwidth,height=!,clip]{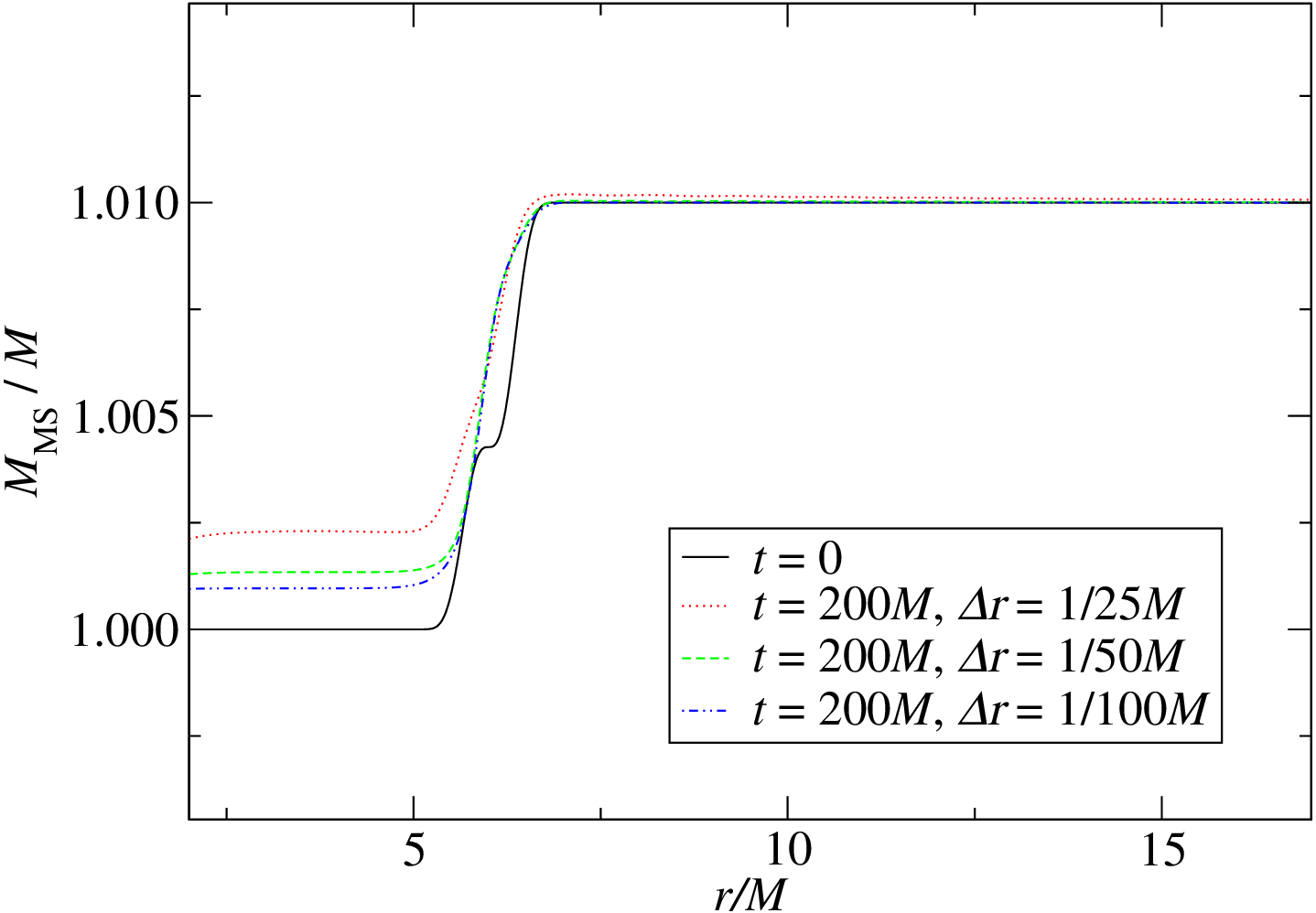}
\caption{As in fig. \ref{mass001st}, we show the mass function, this time for the
non-time-harmonic
  case. The initial mass of the scalar field is $m_{\rm sf}=0.01M$. This time about 10\% of it
  falls into the black hole. \label{mass001} }
\end{figure}

Figures~\ref{mass001st} and~\ref{mass001} show the Misner-Sharp mass
function (equation~\eq{MSdef}) for both types of 
initial data. The continuous line shows the initial value ($M_{\rm
MS}$ at $t=0$). The discontinuous lines show $M_{\rm MS}$ at $t=200M$ for
three different resolutions. In both cases the asymptotic value of the
mass stays constant, indicating no scalar field energy is radiated away.
An inspection of the mass behavior at smaller radii for the solution obtained
with time-harmonic initial data reveals that this converges 
to essentially the initial value, thus  a negligible amount
of mass falls into the black 
hole. For the non-time-harmonic case about $10\%$ of the field's
initial mass falls into the black hole. 

The amount of mass that falls into the black hole is
calculated by subtracting the Misner-Sharp mass at the horizon, minus
the initial mass of the black hole. In the case of time-harmonic initial
data this number is $(1\pm3)\times 10^{-4}M$, while for that of
non-time-harmonic initial data it is $(10\pm3)\times 10^{-4}M$ (see
table~\ref{table:bh_mass} and figure~\ref{bh_mass}). These values are
calculated using the highest resolution ($\Delta r=1/100M$), and the
errors as the difference of these values with those of a lower resolution
($\Delta r=1/50M$).
%
%%% FIG %%% Mass function. Stationary state
%
%%% FIG %%% Mass function. Non-stationary state

%\clearpage
\subsubsection*{Non-linear case}

We turn now to the non-linear cases investigated. These
correspond to initial mass configurations where the scalar
field has a mass of at least 10\% of that of the black hole. In this
regime we solely adopt the ``pulse'' prescription defined
in equation~\eq{pulsedef} for the scalar field since the time-harmonic data
is obtained under an assumption which is no longer valid.

\begin{figure*}[!tbh]
\subfigure[]{
  \includegraphics[angle=0,width=\columnwidth,height=!,clip]{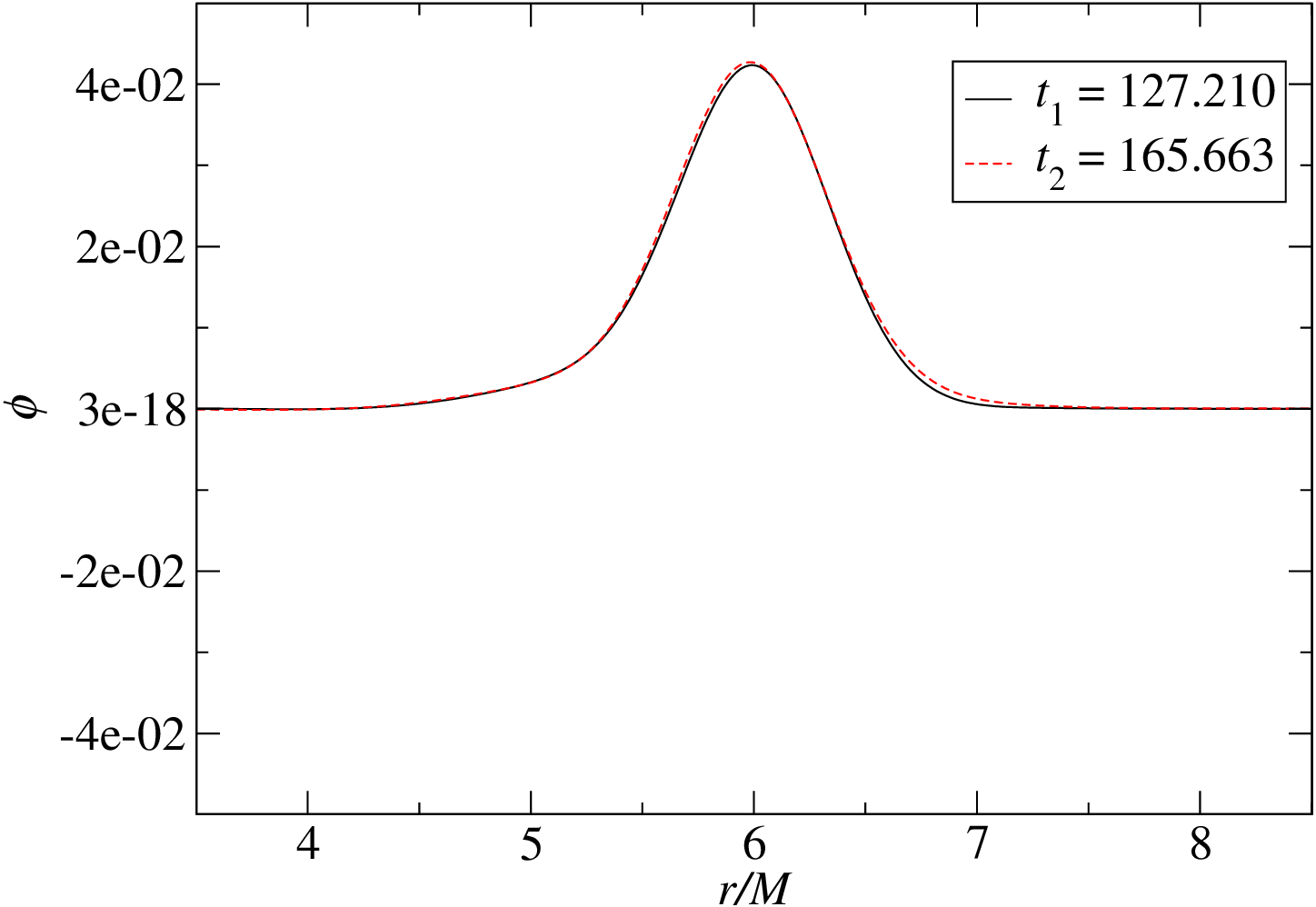}
\label{ns010_a} }
\subfigure[]{
  \includegraphics[angle=0,width=\columnwidth,height=!,clip]{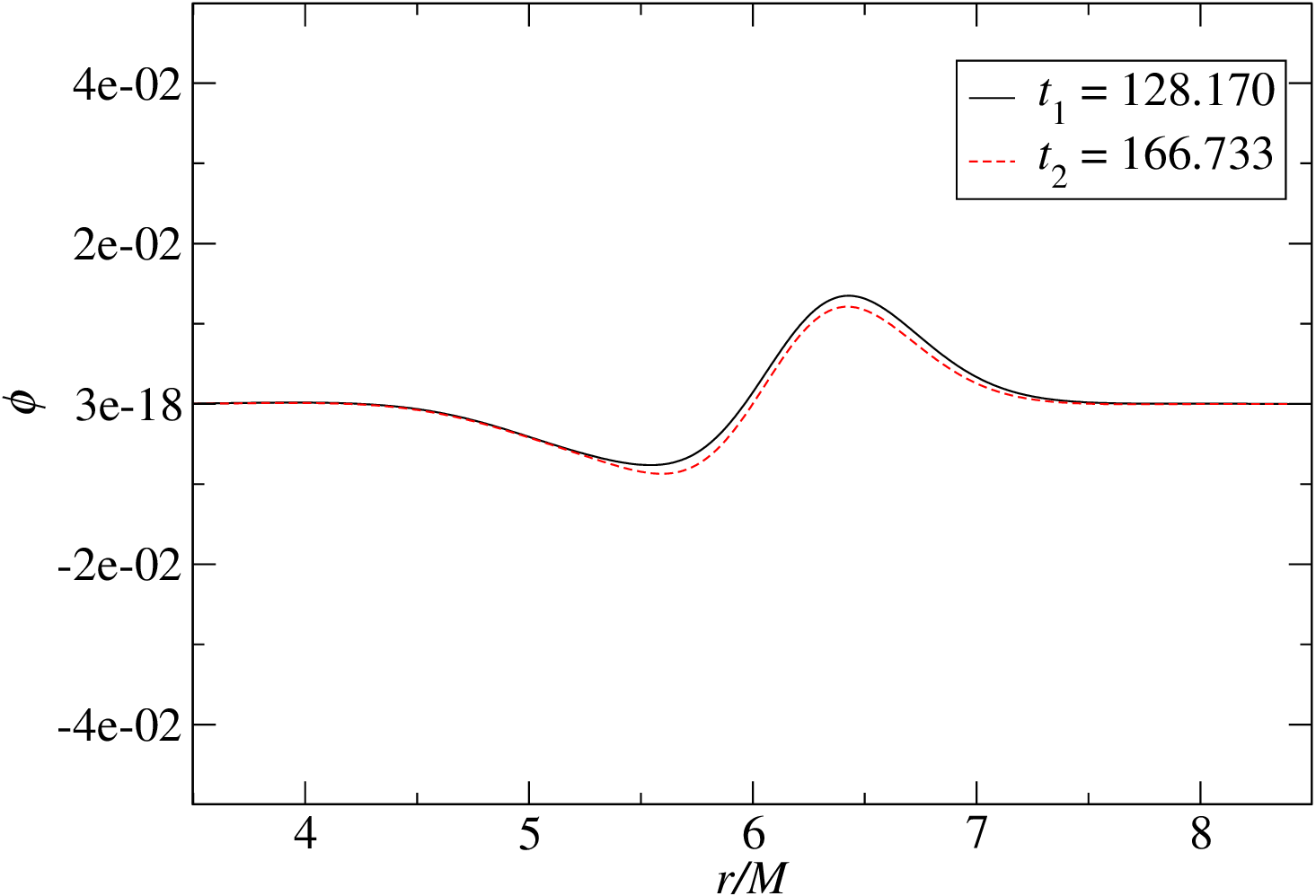}
\label{ns010_b} }
\caption{
The scalar field at different times is compared to check if the
solution obeys a harmonic time dependence. Case with non-time-harmonic
initial data. Initial mass of the  
  scalar field $m_{\rm sf}=0.10M$. Figure~\ref{ns010_a} shows the scalar
  field when it reaches a maximum, while figure~\ref{ns010_b} shows it at
  about a quarter of a period later. In both cases, the profile shown in continuous
  line is separated 10 periods from the one in dashed line.
\label{ns010}} 
\end{figure*}

\begin{figure*}[!tbh]
\subfigure[]{
  \includegraphics[angle=0,width=\columnwidth,height=!,clip]{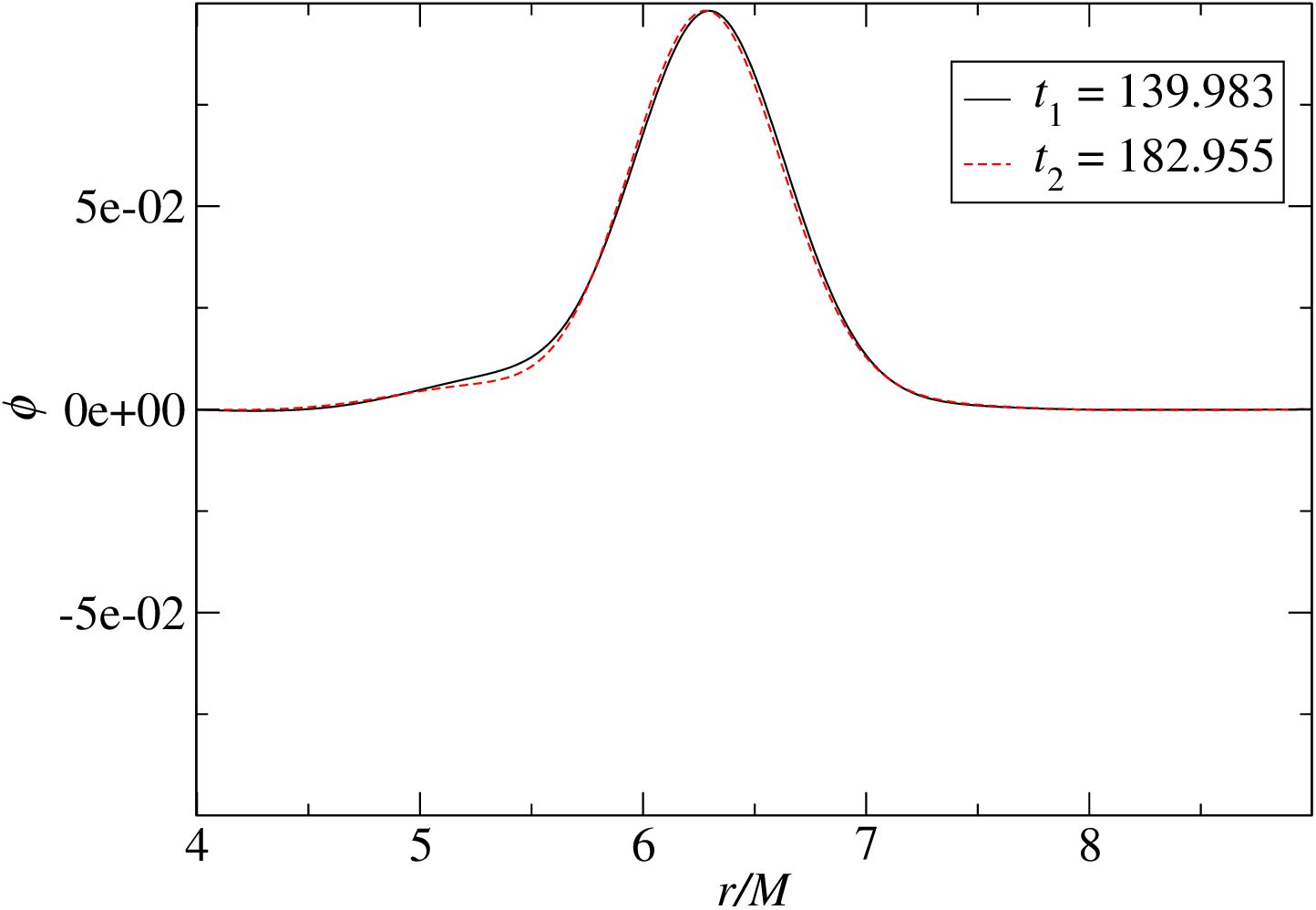}
\label{ns050_a} }
\subfigure[]{
  \includegraphics[angle=0,width=\columnwidth,height=!,clip]{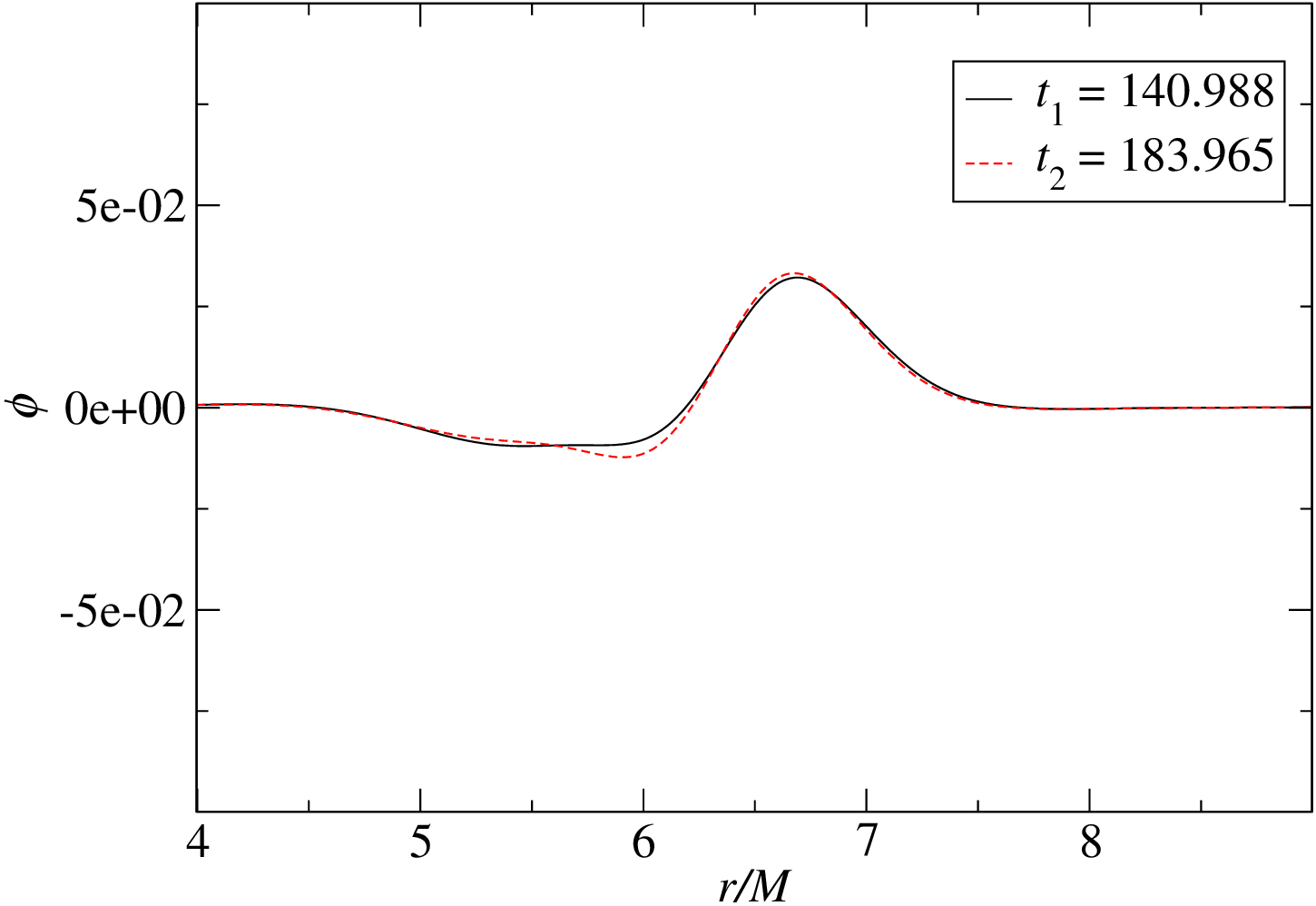}
\label{ns050_b} }
\caption{
This figure shows the same comparisons as figure~\ref{ns010}, but for
an initial mass of the scalar field of $m_{\rm sf}=0.50M$. The
separation between the profiles compared is also 10 periods.
 \label{ns050}}
\end{figure*}

As we have done for the linear case, we also compare profiles at
different times for simulations with higher initial $m_{\rm
  sf}$. Figures~\ref{ns010} and~\ref{ns050} correspond to initial
masses of the scalar field of $m_{\rm sf}=0.10M$ and $m_{\rm
  sf}=0.50M$, respectively. The time it takes to reach a 
state described by a harmonic time dependence is longer than 
in the linear regime, especially for the higher
initial $m_{\rm sf}=0.50M$. For that reason, the first samplings
(labeld $t_1$ in the figures) occur later than in the linear case,
and the interval between the profiles compared, $t_2-t_1$, is ten
periods, as opposed to 22 in the linear cases.

%
%%% FIG %%% Profiles. msf=0.10
%
%%% FIG %%% Profiles. msf=0.50

\begin{figure*}[!tbh]
\subfigure[Initial $m_{\rm sf}=0.10M$]{
  \includegraphics[angle=0,width=\columnwidth,height=!,clip]{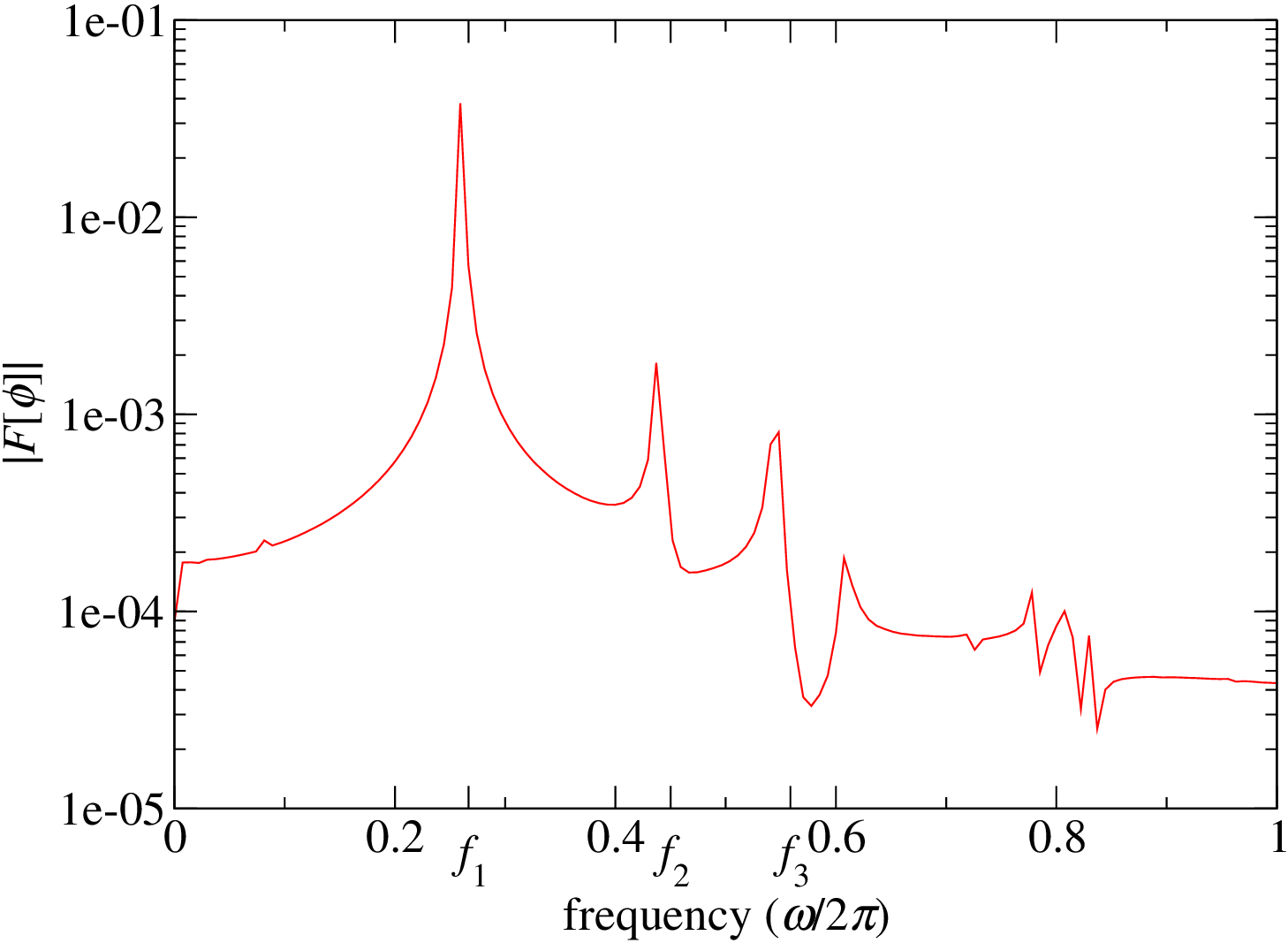}
\label{DFT_ns010} }
\subfigure[Initial $m_{\rm sf}=0.50M$]{
  \includegraphics[angle=0,width=\columnwidth,height=!,clip]{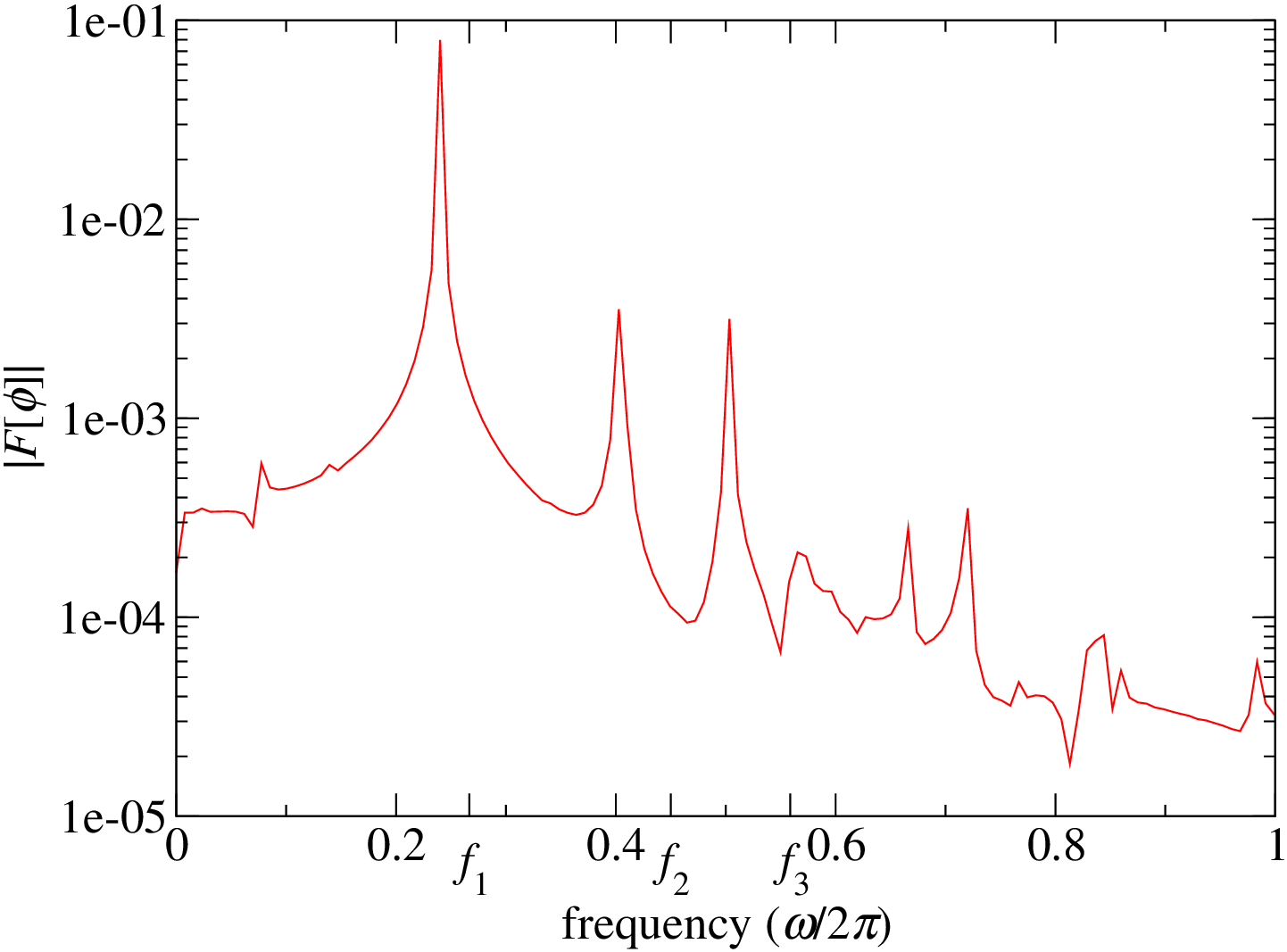}
\label{DFT_ns050} }
\caption{Absolute value of the discrete Fourier transform in $t$ of the space integral $\int \!
  \phi(r,t)\, dr$. The marks labeled $f_n$ denote the frequencies of
  the first modes obtained from the shooting. The three peaks, which indicate
  the dominant frequencies in the solution, lie
  at slightly lower frequencies than those of the time-harmonic states in
  the linear case. This behavior is consistent with the frequency shift due
  to the black hole growing in size. However, the growth alone does not fully
  account for the observed shift, though this is expected as non-trivial
  contribution due to non-linearities also play a role.
\label{DFT_ns}}
\end{figure*}

The absolute value of the Fourier transform of $\int \! \phi\;dr$, $|F[\phi]|$, is shown in
figure~\ref{DFT_ns} for the two different initial masses of $\phi$. 
Again, we compute the transformation after the initial transient
behavior has passed and the scalar filed has already reached a quiescent state. 
As a useful indicator, we also show the frequencies corresponding to time-harmonic states.
Now, while the observed modes do not coincide exactly with those
obtained at the linear approximation, they are
close to them.
%
%%% FIG %%% Fourier transform. All non- cases

\begin{figure*}[!tbh]
\subfigure[Initial $m_{\rm sf}=0.10M$]{
  \includegraphics[angle=0,width=\columnwidth,height=!,clip]{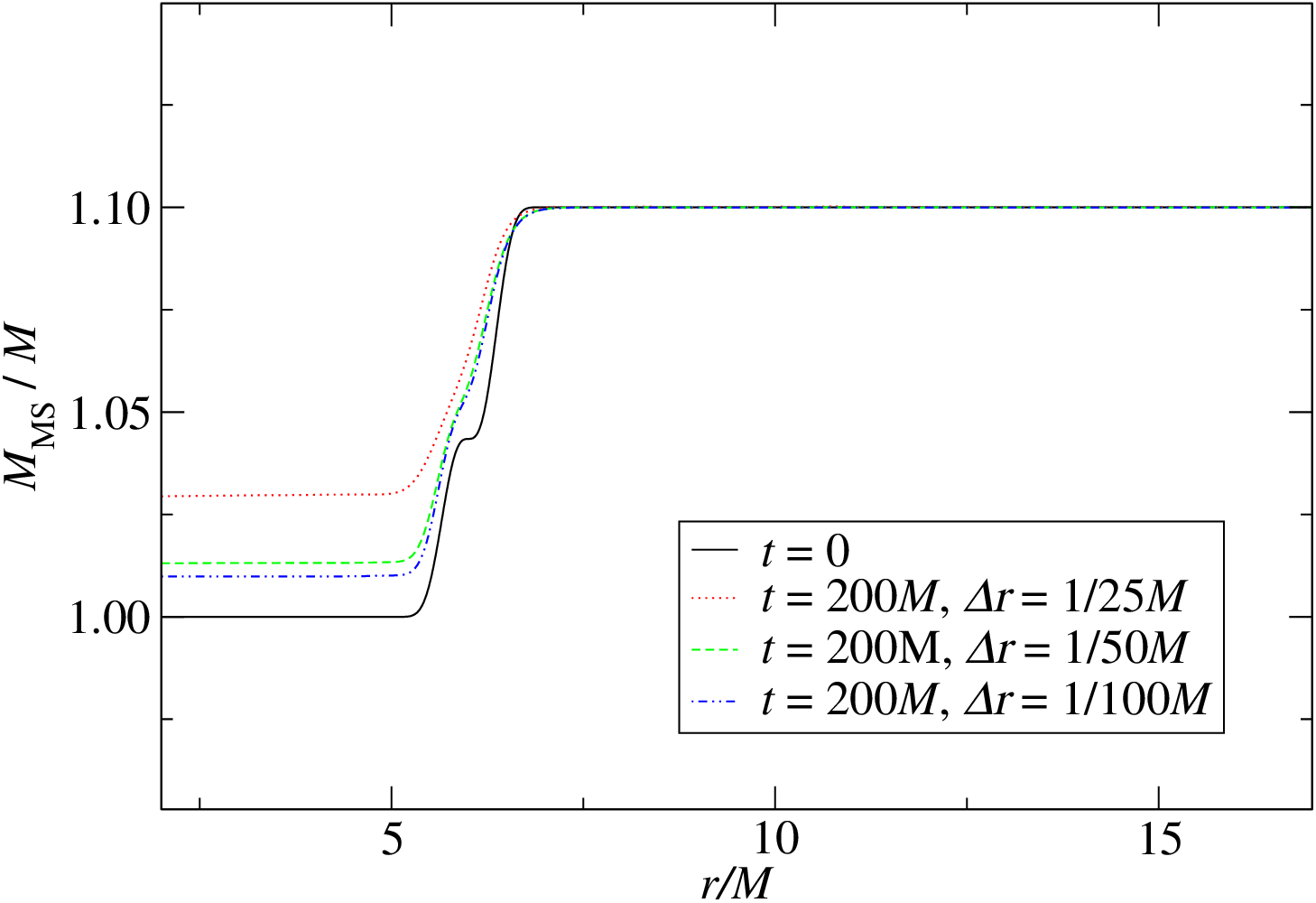}
\label{mass010} }
\subfigure[Initial $m_{\rm sf}=0.50M$]{
  \includegraphics[angle=0,width=\columnwidth,height=!,clip]{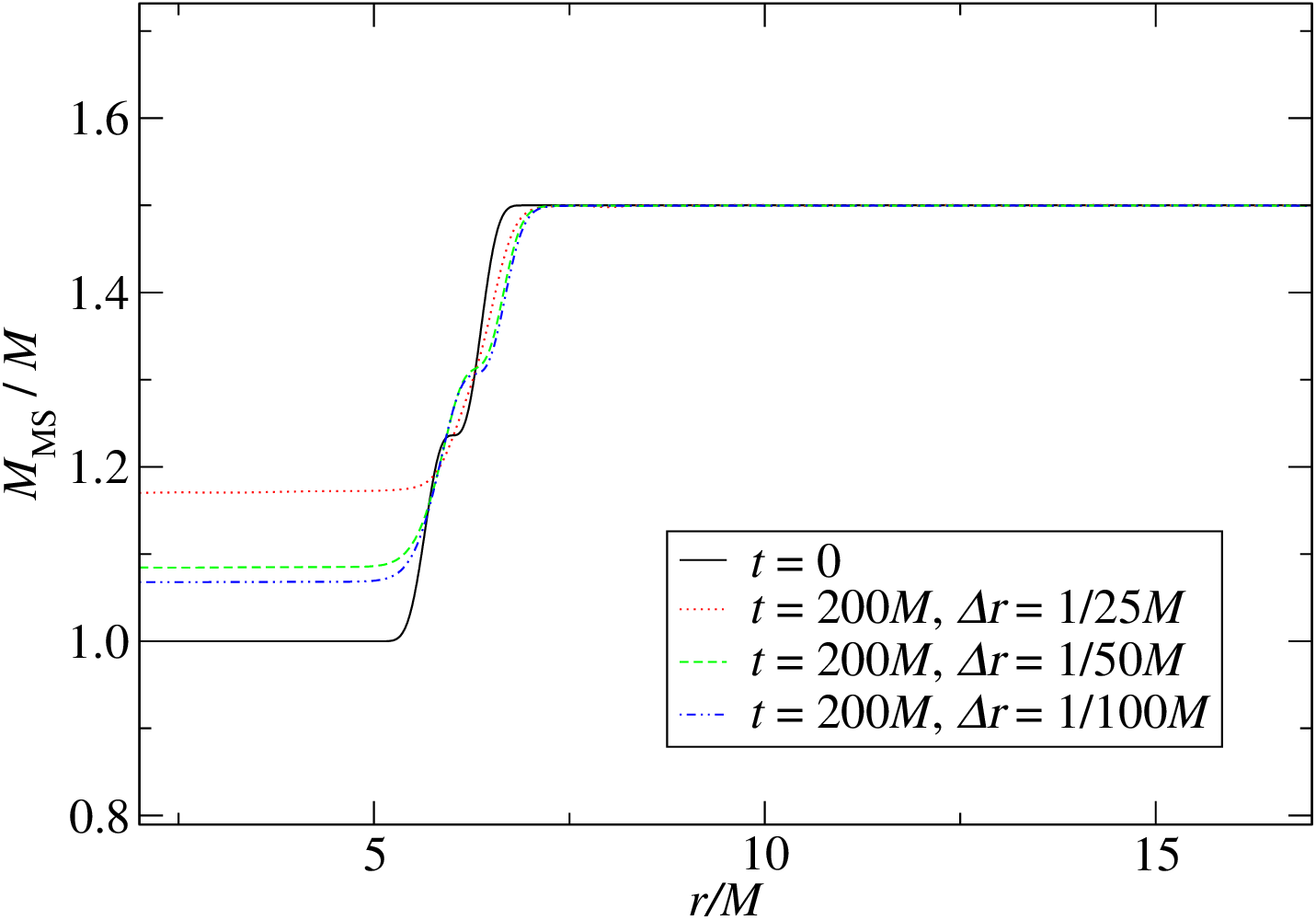}
\label{mass050} }
\caption{Mass function at $t=0$; and at $t=200M$ for three
  resolutions. The discontinuous lines show the
  mass function at t=200M for three resolutions. In
  each of these cases,
  about 10\% of the initial mass of the scalar field falls into the
  black hole, 
  while nothing  escapes to infinity. \label{mass}}
\end{figure*}

\begin{figure}[!tbh]
  \includegraphics[angle=0,width=\columnwidth,height=!,clip]{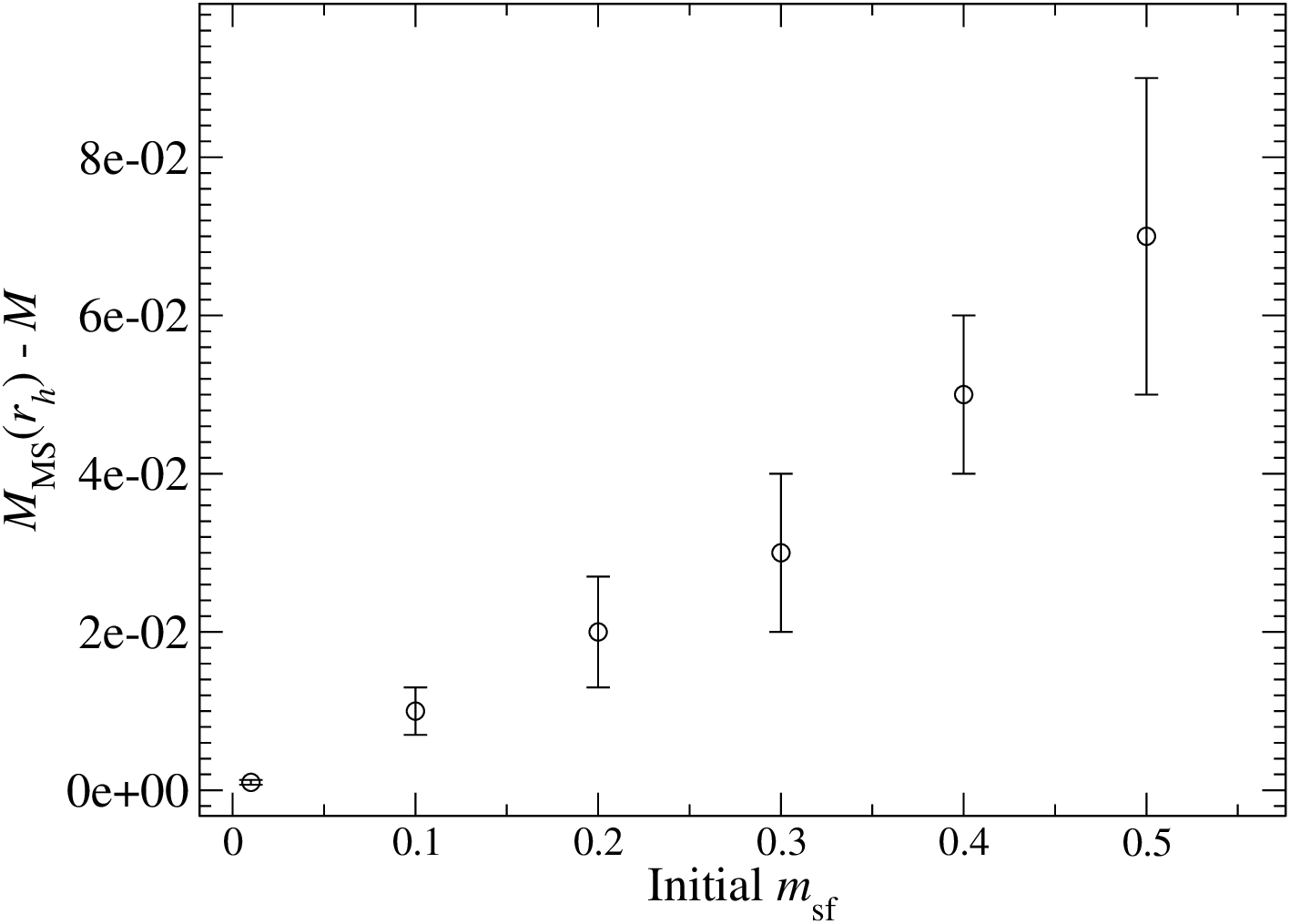}
  \caption{Mass that falls into the black
  hole for different initial masses of the scalar field. Calculated as the
  Misner-Sharp mass at the horizon at $t=200$ 
  minus the initial mass of the black hole. See
  table~\ref{table:bh_mass}. \label{bh_mass} }
\end{figure}

In figure~\ref{mass} we show the Misner-Sharp mass at $t=0$; and at
$t=200M$ for three different resolutions. Figures~\ref{mass010}
and~\ref{mass050} correspond to initial masses of the scalar field of
$m_{\rm sf}=0.10M$ and $m_{\rm sf}=0.50M$, respectively. In
all these cases about 10\% of the scalar field's mass falls into the
black hole, while nothing escapes outwards. Additionally, for
the case with grater mass, the scalar filed spreads slightly outwards
before reaching a quiescent state.
%
%%% FIG %%% Mass function. All non-stat. cases
Although we only show figures corresponding to two different initial
values of $m_{\rm sf}$, we have simulated the system for other values of this
parameter $m_{\rm sf} = \kappa\,10^{-1} M$ ($\kappa=1...5$). In all these cases essentially
no scalar field energy is radiated away, while a small portion falls into the
black hole. The measured values are shown in table~\ref{table:bh_mass} and figure~\ref{bh_mass}.
\begin{table}[!tbh]
\caption{Mass that falls into the black
  hole for different initial masses of the scalar field. Calculated as the
  Misner-Sharp mass at the horizon at $t=200$ 
  minus the initial mass of the black hole. See
  figure~\ref{bh_mass}. \label{table:bh_mass}} 
\begin{ruledtabular}
\begin{tabular}{cc}
%  \hline
  Initial $m_{\rm sf}\;[M]$ & ($M_{\rm MS}(r_{\rm h}) - M) \; [10^{-2}M]$ \\
  \hline 
  $0.01$ & $0.10  \pm 0.03 $ \\
  $0.10$ & $1.0  \pm 0.3 $\\
  $0.20$ & $2.9  \pm 0.7 $\\
  $0.30$ & $3  \pm 1 $\\
  $0.40$ & $5  \pm 1 $\\
  $0.50$ & $7  \pm 2 $\\
%  \hline
\end{tabular}
\end{ruledtabular}
\end{table}

\begin{figure}[!tbh]
  \includegraphics[angle=0,width=\columnwidth,height=!,clip]{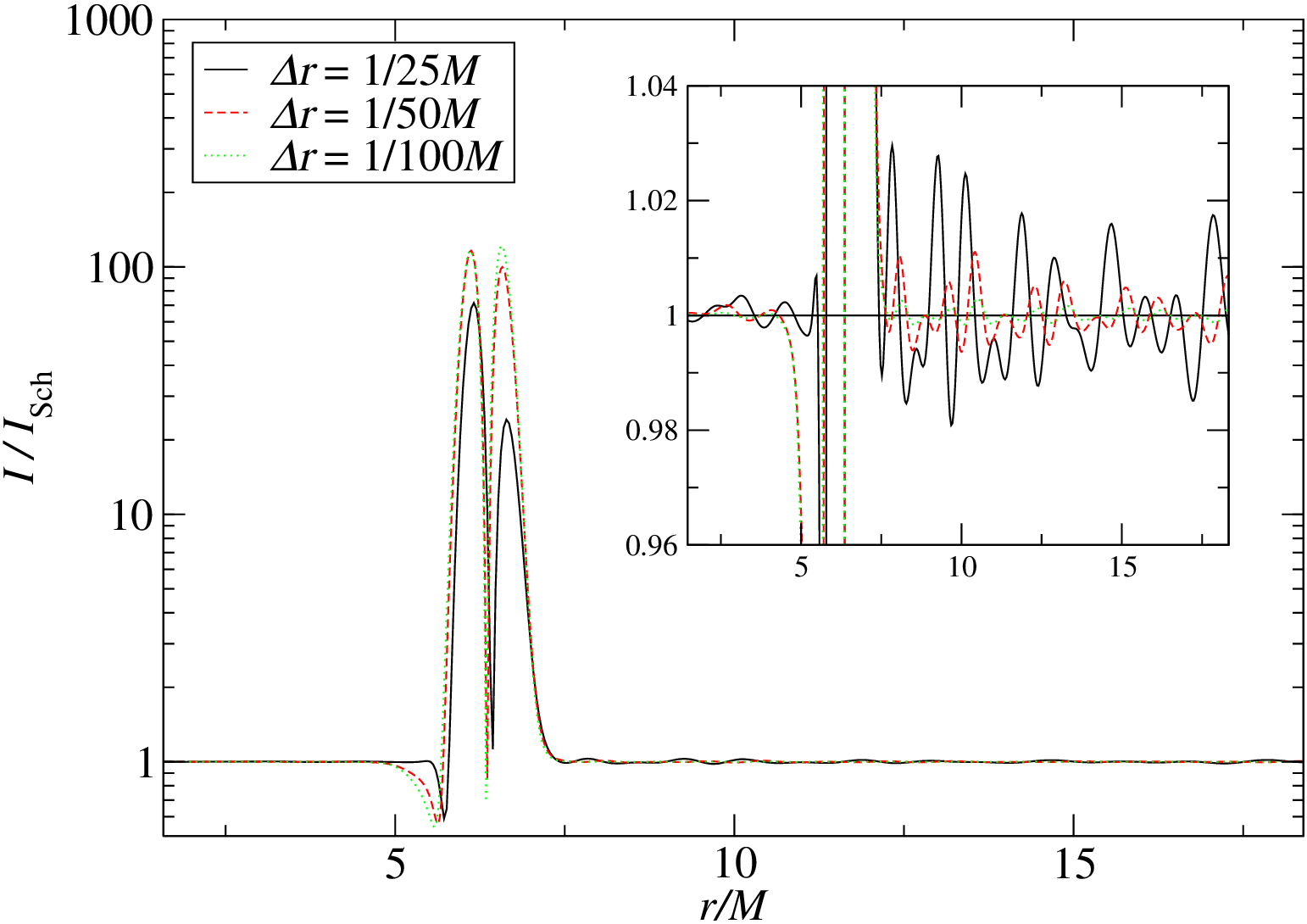}
  \caption{Kretschmann invariant quotient for three resolutions at $t_1=139.983M$. 
 This quotient
  converges to 1 outside of the region where the scalar field is
  confined. A horizontal line at $I/I_{\rm Sch}=1$ have been drawn
  as a guide. Initial $m_{\rm sf}=0.50M$.  \label{RRoverRR} }
\end{figure}

If, after some transient time, the scalar field is finally confined
within a compact region, lets say $[r_a,r_b]$, 
the space-time   should be that of Schwarzschild for $r>r_b$,
with a Schwarzschild mass equal to the total mass inside the sphere
$r=r_b$. This can be checked by evaluating the Kretschmann
invariant. In figure~\ref{RRoverRR} we show the quotient $I/I_{\rm
  Sch}$ (see the paragraph containing equation~\eq{KI}) at
$t\approx140M$ for the case with initial $m_{\rm 
  sf}=0.5M$. This quotient converges to one for
$r>r_b$, and also for $r<r_a$. 

\clearpage

%%%%%%%%%%%%%%%%%%%%%%%%%%%%%%%%%%%%%%%%%%%%%%%%%%
\section{Conclusions} \label{conclusions}
%%%%%%%%%%%%%%%%%%%%%%%%%%%%%%%%%%%%%%%%%%%%%%%%%%
We have discussed difficulties encountered when
attempting to confine a scalar field distribution within
some region. The existence of a symmetry in the spacetime
allows for doing so in a consistent manner. For the specific
spherically symmetric case, we have given prescriptions for 
implementing a scalar field with a potential depending on the 
areal radius $R$. 

We have illustrated the viability of this approach
by confining a scalar field distribution around a black hole.
For our particular choice of potential and initial scalar field, the
scalar field becomes totally confined after some transient time, which depends on
the initial mass. During the transient, part of the scalar field
accretes into the black hole, while nothing escapes to infinity.
By adjusting the depth of the potential, the amount of energy
that falls in can be controlled. 

The approach can be exploited, an extended, to mimic situations
of interest. These can range from physical studies of particular
systems, to serve as a testing model for infrastructure development
aimed to simulate more complex systems.

%%%%%%%%%%%%%%%%%%%%%%%%%%%%%%%%%%%%%%%%%%%%%%%%%%
\acknowledgements
We would like to thank M.~Anderson, D.~Garfinkle, C.~Palenzuela-Luque,
J.~Pullin and R.~Wald for helpful discussions as well as
M. Tiglio and J. Pullin for comments  and suggestions on the manuscript.
This work was supported in part by NSF grants PHY-0244699, PHY-0244335,
PHY0326311 and PHY0554793  to Louisiana State University.
The simulations described here were performed on
local clusters in the Dept. of Physics \& Astronomy.
L.L is grateful to the Alfred P Sloan Foundation and
Research Corporation for financial support.

%%%%%%%%%%%%%%%%%%%%%%%%%%%%%%%%%%%%%%%%%%%%%%%%%%
\appendix

%%%%%%%%%%%%%%%%%%%%%%%%%%%%%%%%%%%%%%%%%%%%%%%%%%
\section{Coordinate-Dependent Potential in Axial Symmetry}  \label{axial}
%%%%%%%%%%%%%%%%%%%%%%%%%%%%%%%%%%%%%%%%%%%%%%%%%%

Following \cite{chandra}, we write the (general) {axi-symmetric} line
element in the form
\bea
 ds^2  &=&  -e^{2\nu} (dt)^2 + e^{2\psi}\left(d\ph -q_1 dx^1 -q_2 dx^2
 -\omega dt\right)^2  \nonumber \\
      &&  +e^{2\mu_1} (dx^1)^2 +e^{2\mu_2} (dx^2)^2,
\eea
where all the functions appearing here are functions of $x^0\equiv t$,
$x^1$, and $x^2$, 
but independent of ${x^3\equiv\ph}$. 
We assume that the scalar field $\phi$ is independent of $\ph$, and
hence use ${H^a}_b$ as given in \eq{one}. Evaluating $\cov_a {H^a}_b$
(and assuming that  $h$ and 
$b$ are independent of
$\ph$) we find  
that the $\ph$-component is of the form ${(h-b)}$ times an
expression depending on the metric functions and their
derivatives. We assume that the expression multiplying $(h-b)$ is not
zero, because at 
this moment we want to consider the  case of no other symmetry other
than the axial symmetry. Then, setting this component to zero, we have
the condition 
${b=h}$, which, as we have seen earlier, implies that $h$ is a
constant. This means that the potential will be independent of the
coordinates.

Consider now the special case of axial symmetry without rotation. We
can write the line element in the form
\be
 ds^2 = g_{\ph\ph}d\ph^2 + g_{ij}dx^i dx^j, \quad i,j\neq3
\ee
Evaluating $\cov_a {H^a}_b$, the $\ph$-component this
time results identically  zero, and, setting the other components to zero,
we have:
\bea   \label{hbt}
      \frac{dg_{\ph\ph} }{ dt\;\;}(h-b)
 +2 g_{\ph\ph}   \frac{dh}{dt}   &=& 0, \\
       \label{hbx1}
       \frac{dg_{\ph\ph} }{ dx^1\;\;}(h-b)
 +2 g_{\ph\ph}   \frac{dh}{dx^1}   &=& 0, \\
       \label{hbx2}
       \frac{dg_{\ph\ph} }{ dx^2\;\;}(h-b)
 +2 g_{\ph\ph}   \frac{dh}{dx^2}   &=& 0.
\eea

At this point, one can follow the same procedures as in section
\ref{sphericalcase} (compare these equations to \eq{hb_t} and
\eq{hb_r}). For that reason, in this section we will just summarize the results.

Equations \eq{hbt}-\eq{hbx2} are satisfied if, (i):  $h$ depends on the
coordinates only through an arbitrary
function of $g_{\ph \ph}$,
\be
h(t,x^1,x^2)=f(g_{\ph \ph}(t,x^1,x^2)),
\ee
and (ii): $b$ is given in terms of $h$ by
\be  \label{sol_b}
 b = h + 2 g_{\ph\ph} \frac{\d h}{\d g_{\ph\ph}}. 
\ee
Given these conditions, one can express $\cov_a{T^a}_b$ with $\cov_b
\phi$ as a common factor, and, setting it to zero, obtain the equation
of motion for the scalar field,
\be
   \cov_a \cov^a \phi +
  \frac{\d f}{\d \phi} \;h(g_{\ph\ph}) = 0 ,
\ee
where $f$ and $h$ are arbitrary functions of $\phi$ and $g_{\ph \ph}$, respectively.
On can, in particular, choose these functions as follows,
\bea
 f(\phi)  &=&  - \half \phi^2 ,\\
 h(g_{\ph\ph})  &=&  m^2 + U({g_{\ph\ph}}) .
\eea
Then, the evolution equation becomes
\be
  \left( \cov_a \cov^a - m^2 - U({g_{\ph\ph}}) \right) \phi = 0,
\ee
where we can interpret 
%the constant $m$ as the mass of the scalar field, and 
$U$ as a coordinate-dependent potential.

%%%%%%%%%%%%%%%%%%%%%%%%%%%%%%%%%%%%%%%%%%%%%%%%%%
\section{The Equations} \label{app_eq}
%%%%%%%%%%%%%%%%%%%%%%%%%%%%%%%%%%%%%%%%%%%%%%%%%%
The equations of motion are
\bw
\bea
\dot{g}_{rr}   &=&   \beta g'_{rr} + 2g_{rr}\beta'-2\aa g_{rr}^{1/2}
g_T K_{rr} \;, \label{grr_dot}\\ 
&& \nonumber \\
%%%%%%%%%%%%%%%%%%%%%
\dot{g}_T   &=&  \beta g'_T -2\aa g_{rr}^{1/2}g_T K_T + \frac{2\beta
g_T}{r}  \; , \label{gT_dot} \\
&& \nonumber \\
%%%%%%%%%%%%%%%%%%%%%
\dot{K}_{rr}   &=&  \beta K'_{rr} - \aa g_{rr}^{-1/2}g_T f'_{rrr} -
\aa'' g_{rr}^{1/2}g_T -6g_T^{-1}g_{rr}^{1/2}\aa f_{rT}^2 +4g_T r^{-1}
g_{rr}^{1/2} \aa' - 6g_T r^{-2}g_{rr}^{1/2}\aa + 2K_{rr}\beta ' \\
&& \nonumber \\
&&   - g_T g_{rr}^{-1/2} \aa K_{rr}^2 +
2g_{rr}^{1/2} \aa K_{rr}K_T - 8g_{rr}^{-1/2} \aa f_{rT}f_{rrr} + 2g_T
g_{rr}^{-3/2} \aa f_{rrr}^2 \nonumber \\
&& \nonumber \\
&&  + 2g_T r^{-1}g_{rr}^{-1/2}\aa f_{rrr} - g_T g_{rr}^{-1/2} \aa f_{rrr}
+ g_T g_{rr}^{1/2} \aa \,4\pi\! \left(T g_{rr}-2 S_{rr}\right)
\nonumber \; , \label{Krr_dot} \\
&& \nonumber \\
%%%%%%%%%%%%%%%%%%%%%
\dot{K}_{T}   &=&   \beta K'_{T} - \aa g_T g_{rr}^{-1/2} f'_{rT} +2
\beta r^{-1}K_T +g_T r^{-2} g_{rr}^{1/2} \aa + \aa g_{T} K_T K_{rr}
g_{rr}^{-1/2} - g_T f_{rT} \aa' g_{rr}^{-1/2} -2 \aa f_{rT}^2 g_{rr}^{-1/2} 
\\ && \nonumber \\
&&  + \aa g_{rr}^{1/2} g_T \, 4\pi \!\left(Tg_T-2S_T\right)
 \label{KT_dot} \; , \\
&& \nonumber \\
%%%%%%%%%%%%%%%%%%%%%
\dot{f}_{rrr}   &=&   \beta f'_{rrr} - \aa g_{rr}^{1/2} g_T K'_{rr}
- 4 g_{rr}^{3/2} \aa' K_T + 12 g_T^{-1} g_{rr}^{3/2}\aa K_T f_{rT} -
4g_{rr}^{1/2}\aa K_T f_{rrr} - g_T g_{rr}^{-1/2}\aa K_{rr}f_{rrr} \\
&& \nonumber \\
&&  -10g_{rr}^{1/2}\aa K_{rr}f_{rT} + 3f_{rrr}\beta' + g_{rr}\beta'' -
\aa' g_{rr}^{1/2} g_T K_{rr} +2 r^{-1} g_T g_{rr}^{1/2} \aa K_{rr} + 8
r^{-1} g_{rr}^{3/2}\aa K_T + 4 \aa g_{rr}^{3/2} g_T \,4\pi\!J_r \nonumber
\; ,\label{frrr_dot}\\
&& \nonumber \\
%%%%%%%%%%%%%%%%%%%%%
\dot{f}_{rT}   &=&   \beta f'_{rT} - \aa g_{rr}^{1/2} g_T K'_T +
 \beta' f_{rT} - \aa' g_{rr}^{1/2} g_T K_T + 2g^{1/2}\aa K_T f_{rT} -
 \aa g_{rr}^{-1/2} K_T f_{rrr} g_T + 2 r^{-1} \beta f_{rT} 
 \; ,\label{frT_dot} \\
&& \nonumber \\
%%%%%%%%%%%%%%%%%%%%%%
\dot{\Phi }   &=&    \beta \Phi ' - \aa g_{rr}^{1/2}g_T \Pi' -
 g_{rr}^{-1/2} \aa g_T \Pi f_{rrr} + 2\aa g_{rr}^{1/2} \Pi f_{rT} + 2
 r^{-1} \aa g_{rr}^{1/2} g_T \Pi - \aa' g_{rr}^{1/2} g_T \Pi + \Phi \beta'
 \;, \label{Phidot} \\
&& \nonumber \\
%%%%%%%%%%%%%%%%%%%%%%
\dot{\Pi }   &=&   \beta \Pi ' - g_{rr}^{-1/2} \aa g_T \Phi' +
 g_{rr}^{-1/2} \aa g_T \Pi K_{rr} + 2\aa g_{rr}^{1/2} \Pi K_T -4
 g_{rr}^{-1/2} \aa \Phi f_{rT} + 2r^{-1} g_{rr}^{-1/2} \aa g_T \Phi -
 g_{rr}^{-1/2} g_T \Phi \aa' \\ && \nonumber \\
&&  + g_{rr}^{1/2}g_T\aa V \phi
 \;,  \label{Pidot} \\
&& \nonumber \\
%%%%%%%%%%%%%%%%%%%%%%
\dot{\phi}  &=&  \beta \phi' - g_T g_{rr}^{1/2} \aa
\Pi\;,\label{phidot} 
\eea
\ew
where $\tilde{\alpha}=\alpha r^2 \sin\theta=N/\sqrt{g_{rr}}g_T$; dots
denote derivative with respect to $t$ and primes denote 
derivatives with respect to $r$; and the ``source terms'' are defined in
equations~\eq{rho}-\eq{TST}. 

The constraint equations are
\bea
C  &=&  \frac{f'_{rT}}{g_{rr}g_T} - \frac{1}{2r^2g_T} + 
\frac{f_{rT}\left(\frac{2}{r}+\frac{7f_{rT}}{2g_T}
- \frac{f_{rrr}}{g_{rr}}\right)}{g_{rr}g_T}  +\nonumber \\
&&- \frac{K_T\left(\frac{K_{rr}}{g_{rr}}+\frac{K_T}{2g_T}\right)}{g_T} 
+ 4\pi\rho \;, \; \label{C} \\ 
&& \nonumber \\
%%%%%%%%%%%%%%%%%%%%%%
C_r  &=&  \frac{K'_T}{g_T} + \frac{2K_T}{rg_T} -
\frac{f_{rT}\left(\frac{K_{rr}}{g_{rr}}+\frac{K_T}{g_T}\right)}{g_T}
+\nonumber \\
&&+ 4\pi J_r \;, \label{Cr} \\
%%%%%%%%%%%%%%%%%%%%%%
C_{rrr}  &=&  g'_{rr} + \frac{8g_{rr}f_{rT}}{g_T} - 2f_{rrr} \;,
\label{Crrr} \\
&& \nonumber \\
%%%%%%%%%%%%%%%%%%%%%%
C_{rT}  &=&  g'_T + \frac{2g_T}{r} - 2f_{rT} \;, \label{CrT} \\
&& \nonumber \\
%%%%%%%%%%%%%%%%%%%%%%
C_m  &=&  \Phi - \phi' \;, \label{Cm}
\eea
where the ``source'' terms are difined as
\bea
4\pi \rho  &=&  \frac{V\phi^2}{2} + \frac{\Phi^2}{2g_{rr}} +
\frac{\Pi^2}{2} \;, \label{rho} \\
&& \nonumber \\
%%%%%%%%%%%%%%%%%%%%%%
4\pi T  &=&  -2V\phi^2 -\frac{\Phi^2}{g_{rr}} + \Pi^2 +\nonumber \\
&&-\phi^2r^2g_T \frac{\partial V}{\partial g_\Omega} \;, \label{T} \\
&& \nonumber \\
%%%%%%%%%%%%%%%%%%%%%%
4\pi J_r  &=&  \Phi \Pi \;, \label{Jr} \\
&& \nonumber \\
%%%%%%%%%%%%%%%%%%%%%%
4\pi \left(Tg_{rr}-2S_{rr}\right)  &=&  -V\phi^2g_{rr} - 2\Phi^2
+\nonumber\\
&&-
\phi^2g_{rr}r^2g_T \frac{\partial V}{\partial g_\Omega} \;, \label{TSrr} \\
&& \nonumber \\
%%%%%%%%%%%%%%%%%%%%%%
4\pi \left(Tg_T-2S_T\right)  &=&  -g_T V \phi^2 \;, \label{TST}
\eea
where $g_{\Omega}=r^2g_T$ (see section~\ref{sphericalcase}).

%%%%%%%%%%%%%%%%%%%%%%%%%%%%%%%%%%%%%%%%%%%%%%%%%%
\section{Characteristic structure} \label{characteristic}
%%%%%%%%%%%%%%%%%%%%%%%%%%%%%%%%%%%%%%%%%%%%%%%%%%
The characteristic modes and eigenvalues obtained at a surface $r=const$ are
given by
\be
\begin{array}{ll} 
u_1 = g_{rr} ,                \quad  &\lambda_1 = \beta  ,\\  
u_2 = g_T ,                   \quad  &\lambda_2 = \beta  ,\\  
u_3 = K_{rr}-f_{rrr}/g_{rr} , \quad  &\lambda_3 = \beta+\aa g_T  ,\\  
u_4 = K_T-f_{rT}/g_{rr} ,     \quad  &\lambda_4 = \beta+\aa g_T  ,\\  
u_5 = K_{rr}+f_{rrr}/g_{rr} , \quad  &\lambda_5 = \beta-\aa g_T  ,\\  
u_6 = K_T+f_{rT}/g_{rr} ,     \quad  &\lambda_6 = \beta-\aa g_T  ,\\  
u_7 = \Pi+\Phi/g_{rr} ,       \quad  &\lambda_7 = \beta-\aa g_T  ,\\  
u_8 = \Pi-\Phi/g_{rr} ,       \quad  &\lambda_8 = \beta+\aa g_T  ,\\  
u_9 = \phi ,                  \quad  &\lambda_9 = \beta  .  
\end{array}
\ee

%%%%%%%%%%%%%%%%%%%%%%%%%%%%%%%%%%%%%%%%%%%%%%%%%%
\section{Code Tests} \label{tests}
\begin{figure}[!tbh]%[!tbh]
  \includegraphics[angle=0,width=\columnwidth,height=!,clip]{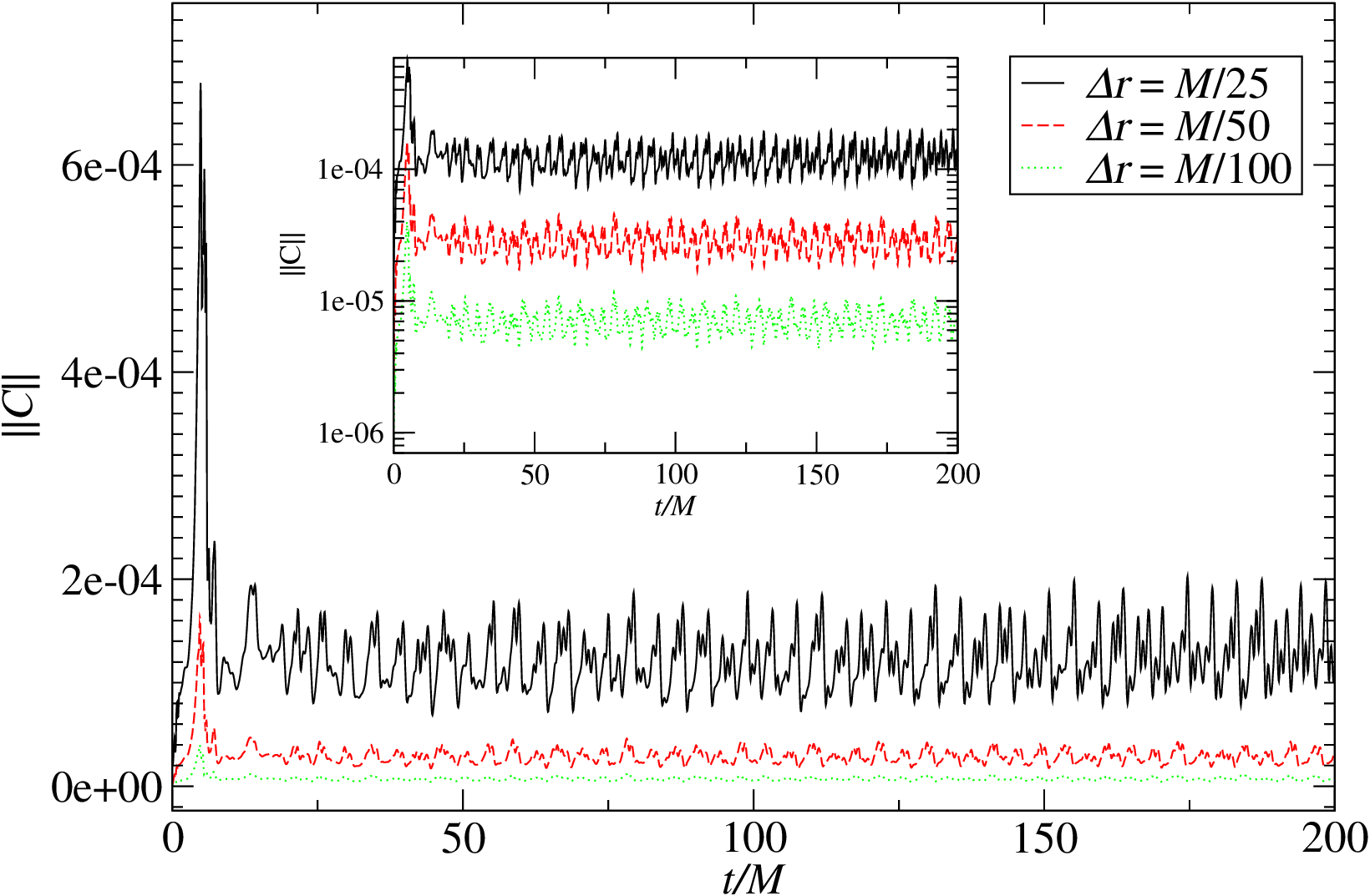}
  \caption{$L_2$ norm of the Hamiltonian constraint (equation \eq{C}) for three
  different resolutions. The measured convergence results of order two as
  expected. \label{plotC} }
\end{figure}

%%%%%%%%%%%%%%%%%%%%%%%%%%%%%%%%%%%%%%%%%%%%%%%%%%
The standard code tests have been performed, showing that all the
constraints and residuals converges to zero with order two. In
figure~\ref{plotC} we show the Hamiltonian constraint, in the case of the
strongest scalar filed studied, that with
initial $m_{\rm sf}=0.5M$. 

The evaluation of the constraints is a particularly important test in
this work, to ensure that the implementation of a coordinate dependent
potential is not breaking the covariance of the theory.

%-------------------------------------------------

%%%%%%%%%%%%%%%%%%%%%%%%%%%%%%%%%%%%%%%%%%%%%%%%%%


\begin{thebibliography}{999}
%%%%%%%%%%%%%%%%%%%%%%%%%%%%%%%%%%%%%%%%%%%%%%%%%%
\bibitem{MTW}
  Misner~C~W, Thorne~K~S and Wheeler~J~A 1973 {\it Gravitation},
  (New York: W~H~Freeman and Company)
\bibitem{scheelteuk}
  Scheel~M~A, Shapiro~S~L and Teukolsky~S~A
  %``Scalar gravitation: A Laboratory for numerical relativity. 3:
  %Axisymmetry,''
 1994 {\it Phys. Rev.} D {\bf 49} 1894
% Boson Star Initial Data
\bibitem{kaup} Kaup~D~J %Klein-Gordon Geon
   1968 {\it Phys. Rev.} {\bf 172} 1331
% Boson Star name is coined (through a HF approximation of the Quantum Description)
\bibitem{ruffini} Ruffini~R and Bonazzola S
     1969  {\it Phys.~Rev.} {\bf 187} 1767 .
     
\bibitem{seidelsuen}
  E.~Seidel and W.~M.~Suen,
  %``Oscillating soliton stars,''
 {\it Phys.\ Rev.\ Lett.\ }  {\bf 66}, 1659 (1991).     
     
\bibitem{liddle}
  Liddle~A~R
  %``An introduction to cosmological inflation,''
  ({\it Preprint} astro-ph/9901124).

\bibitem{Lidsey:1995np}
  J.~E.~Lidsey, A.~R.~Liddle, E.~W.~Kolb, E.~J.~Copeland, T.~Barreiro and M.~Abney,
  %``Reconstructing the inflaton potential: An overview,''
  Rev.\ Mod.\ Phys.\  {\bf 69}, 373 (1997)
  
\bibitem{choptuik}
  Choptuik~M~W
  %``Universality And Scaling In Gravitational Collapse Of A Massless Scalar
  %Field,''
  1993 {\it Phys. Rev. Lett.} {\bf 70} 9.

  
\bibitem{Gundlach:1997wm}
  C.~Gundlach,
  %``Critical phenomena in gravitational collapse,''
  Adv.\ Theor.\ Math.\ Phys.\  {\bf 2}, 1 (1998)
  [arXiv:gr-qc/9712084].
   
\bibitem{bound}
 Peres~A %{\it Absence of Bound States in a Gravitational Field},
 1960 {\it Phys. Rev.} {\bf 120} 1044
\bibitem{bound2}
 Everson~B~L and Brill~D~R 1967 
 %{\em Bound States in the Schwarzschild-Kruskal Geometry}
 {\it Bull. Am. Phys. Soc.} {\bf 12} 578
\bibitem{rosen}
  Rosen~N 1940 {\it Phys.~Rev.} {\bf 147} 150    
\bibitem{kuchar}
  Kuchar~K~V and Torre~C~G 1991
  {\it Phys.~Rev.} D {\bf 44} 3116

\bibitem{frans}
  Choptuik~M~W, Hirschmann~E~W, Liebling~S~L and Pretorius~F 2004 
  {\it Phys. Rev. Lett.} {\bf 93} 131101
\bibitem{unpub}
  Olabarrieta I, Ventrella J, Choptuik M and Unruh W (Unpublished)
  {\it Critical Bahavior in the Gravitational Collapse of a Scalar
  Field with Angular Momentum in Spherical Symmetry}

\bibitem{wald}
 Wald~R~M 1984 {\em General Relativity} (Chicago: The University of Chicago Press)
\bibitem{york_fixing}
 Anderson~A and York~J~W,~Jr. 1999 %{\em Fixing Einstein's Equations},
 {\it Phys. Rev. Lett.} {\bf 82} 4384
\bibitem{Kidder}  
  Kidder~L~E, Scheel~M~A and Teukolsky~S~A 2000
  %{\em  Black hole evolution by spectral methods}, 
  {\it Phys. Rev.} D {\bf 62} 084032
\bibitem{cpbc}
  Calabrese~G, Lehner~L and Tiglio~M %{\em Constraint-preserving
  %boundary conditions in numerical relativity}, 
  2002 {\it Phys. Rev.} D {\bf 65} 104031
\bibitem{nrf}
 Press~W, Flannery~B, Teukolsky~S and Vetterling~W 1992 {\it Numerical
 Recipes in Fortran} (Cambridge University Press)
\bibitem{lsode} Hindmarsh~A~C (Lawrence Livermore National Laboratory,
  http://www.llnl.gov/casc/odepack/)
\bibitem{KS1} Kreiss~H~O and Scherer~G 1974 in {\em Mathematical Aspects
  of Finite Elements in Partial Differential Equations}, edited by
  C D Boor (New York: Academica Press)
\bibitem{KS2} Kreiss~H~O and Scherer~G 1977 {\it Tech. Rep.}  Dept. of
  Scientific Computing, Uppsala University
\bibitem{strand}
  Strand~B,
  %``Summation by parts for finite difference approximations for d/dx'',
  1994 {\it Journal of Computational Physics} {\bf 110} 47
\bibitem{SBP0}
  Calabrese~G, Lehner~L, Neilsen~D, Pullin~J, Reula~O, Sarbach~O and Tiglio~M
  %``Novel finite-differencing techniques for numerical relativity: application
  %to black hole excision,''
  2003  {\it Class. Quant. Grav.}  {\bf 20} L245
\bibitem{SBP1}
  Calabrese~G, Lehner~L, Reula~O, Sarbach~O and Tiglio~M
  %``Summation by parts and dissipation for domains with excised regions,''
  2004 {\it Class. Quant. Grav.} {\bf 21} 5735
\bibitem{tadmor}
  Levy~D and Tadmor~E 1998
  {\it SIAM Journal on Num. Anal.} {\bf 40} 40
\bibitem{KO}
 Kreiss~H and Oliger~J 1973 {\it Methods for the approximate solution of
 time independent problems} (Geneva: GARP Publication Series)
\bibitem{gustaffsonkreissoliger} Gustafsson~B, Kreiss~H and Oliger~J
  1995 {\it Time dependent problems and difference methods}
  (John Wiley and Sons)
\bibitem{SBP2}
  Lehner~L, Neilsen~D, Reula~O and Tiglio~M
  %``The discrete energy method in numerical relativity: Towards long-term
  %stability,''
  2004 {\it Class. Quant. Grav.} {\bf 21} 5819
\bibitem{olsson}
 Olsson~P 1995 {\it Math. Comp.} {\bf 64} 1035; {\bf 64} S23; {\bf 64} 1473 
\bibitem{thomas}
  Thomas~J~W 1995 {\it Numerical Partial Differential Equations: Finite
  Difference Methods}, Texts in Applied Mathematics 22
  (New York: Springer-Verlag)
\bibitem{chandra}
 Chandrasekhar~S 1992 {\it The Mathematical Theory of Black Holes}
 (New York: Oxford University Press) 

\end{thebibliography}
\end{document}